\documentclass{aa}  

\usepackage{graphicx}
\usepackage{txfonts}
\begin{document} 

   \title{Environmental dependence of AGN activity and star formation in galaxy clusters from Magneticum simulations}


   \author{G. Rihtaršič
          \inst{1,2}
          \and
          V. Biffi\inst{3}
          \and
          D. Fabjan\inst{2,3}
          \and
          K. Dolag\inst{1,4}
          }

    \institute{
    Universit\"ats-Sternwarte, Fakult\"at f\"ur Physik, Ludwig-Maximilians-Universit\"at M\"unchen, Scheinerstr. 1, 81679 M\"unchen, Germany 
    \and
    Faculty of Mathematics and Physics, University of Ljubljana, Jadranska 19, 1000 Ljubljana, Slovenia\\
    \email{gregor.rihtarsic@fmf.uni-lj.si}
    \and
    INAF --- Osservatorio Astronomico di Trieste, via Tiepolo 11, 34143 Trieste, Italy
    \and
    Max-Planck-Institut für Astrophysik, Karl-Schwarzschild-Straße 1, 85741 Garching, Germany   
              }

    \date{Received 12 July 2023, accepted 28 December 2023}

  \abstract
   {Environment inside and on the outskirts of galaxy clusters has a profound impact on the star formation rate and AGN activity in cluster galaxies. While the overall star formation and AGN suppression in the inner cluster regions has been thoroughly studied in the past, recent X-ray studies also indicate that conditions on the cluster outskirts may promote AGN activity.  }
   {We investigate how the environment and the properties of host galaxies impact the levels of AGN activity and star formation in galaxy clusters. We aim to identify significant trends in different galaxy populations and suggest possible explanations.}
   {We studied galaxies with stellar mass $\log M_* (M_\odot)>10.15$ in galaxy clusters with mass $M_{500}>10^{13}M_\odot$ extracted from box2b (640 comoving Mpc/$h$) of the {\it Magneticum Pathfinder} suite of cosmological hydrodynamical simulations at redshifts 0.25 and 0.90. We examined the influence of stellar mass, distance to the nearest neighbouring galaxy, clustercentric radius, substructure membership and large-scale surroundings on the fraction of galaxies hosting an AGN, star formation rate and the ratio between star-forming and quiescent galaxies. }
   {We found that in low-mass galaxies, AGN activity and star formation are similarly affected by the environment and decline towards the cluster centre. In massive galaxies, the impact is different; star-formation level increases in the inner regions and peaks between 0.5 and 1 $R_{500}$ with a rapid decline in the centre, whereas AGN activity declines in the inner regions and rapidly rises below $R_{500}$ towards the centre. We suggest that this increase results from larger black hole masses relative to stellar masses in the cluster centre. After disentangling the contributions of neighbouring cluster regions, we found an excess of AGN activity in massive galaxies on the cluster outskirts ($\sim 3 R_{500}$). We also found that the local density, substructure membership and stellar mass strongly influence star formation and AGN activity but verified that they cannot fully account for the observed radial trends.}
   {}


\titlerunning{AGNs and star formation in galaxy clusters from Magneticum simulations}
\authorrunning{G. Rihtaršič et al.}
\maketitle

\section{Introduction}
\label{introduction}
Galaxy clusters are the largest collapsed structures in the Universe. Their deep gravitational potential wells, densely populated with galaxies and filled with hot cluster gas, host extreme conditions that can significantly influence the evolution of cluster galaxies. In particular, different processes govern the abundance and distribution of cold gas, which fuels star formation and active galactic nuclei (AGNs) in cluster galaxies. We briefly summarize those processes in the following.

As the galaxies are moving through the intracluster medium, they are exposed to ram pressure which can remove the gas from the galaxy in the process called \textit{ram-pressure stripping} \citep[e.g.,][]{2022A&ARv..30....3B, 2016A&A...587A..68B, 2000Sci...288.1617Q, 1999MNRAS.308..947A, 1972ApJ...176....1G}. Additionally, the cluster environment may prevent the accretion of the surrounding gas onto cluster galaxies, causing the cold gas reservoir to be gradually depleted by star formation in the process called \textit{strangulation} \citep[e.g.,][]{2015Natur.521..192P, 1980ApJ...237..692L}. The processes operate on different timescales; Whereas ram pressure stripping is expected to quench
the galaxy abruptly after the infall, strangulation is a longer and
more gradual process. However, both result in a larger abundance of early-type morphologies (see \citealt{2023arXiv230202376V} and references therein), lower levels of star formation \citep[e.g.,][]{ 2014ApJ...783..136C, 
2012MNRAS.424..232W, 
2010MNRAS.404.1231V, 
2006A&A...446..839G, 
1998ApJ...504L..75B, 
1983AJ.....88..483K} and suppressed AGN activity \citep{2018A&A...620A..20K, 2014MNRAS.437.1942E,  
 2007ApJ...664..761M, 2010ApJ...714L.181K,
 2012ApJ...754...97H} in galaxy clusters compared to the field population of galaxies. The suppression is stronger in more massive clusters \citep[][]{2018A&A...620A..20K, 2015MNRAS.446.2709E}. 

The suppression of star formation due to environmental effects (i.e., environmental quenching) is particularly significant in the evolution of low-mass galaxies that still contain enough cold gas prior to infall. Massive galaxies, on the other hand, can be quenched on their own in isolation; they succumb to the so-called mass quenching (e.g., see \citealt{2010ApJ...721..193P}, \citealt{2016MNRAS.462.4495H}, \citealt{2017MNRAS.469.3670S}, \citealt{2019ApJ...878...69L}; see also \citealt{2019MNRAS.488.5370L} and \citealt{2017MNRAS.472.4769T}, for results from the \textit{Magneticum} simulations). Several causes for mass quenching have been proposed, such as the feedback processes from supernovae and local ionisation \citep[e.g.,][]{2010MNRAS.403L..16C}. Furthermore, the AGN activity in massive galaxies itself can lead to ejection and heating of the gas and can terminate star formation \citep[][]{2015MNRAS.448.1504S,2015ARA&A..53..115K, 2012ARA&A..50..455F}. For that reason, the connection between star formation and AGN activity is still debatable, even though both are fueled by the same cold gas reservoir in the host galaxy.
The correlation between star formation and nuclear activity has been confirmed in several studies \citep[e.g.,][]{2020MNRAS.497.3273F,  2012ApJ...753L..30M, 2012ApJ...746..168D}, while some studies favour the quenching role of AGNs \citep[e.g.,][]{2016MNRAS.455L..82L, 2007ApJ...660L..11N}.

Despite the net quenching influence of the cluster environment, processes, such as ram-pressure stripping, can, in some cases, promote star formation and AGN activity: while the hot gas can be removed from galaxies during their first infall into the cluster, some cold and dense clouds may remain bound within the galactic potential. When approaching the increasingly pressurized cluster centre, the pressure (ram pressure and static thermal pressure) may compress the remaining gas, leading to a short period of enhanced star formation, reported in several numerical studies \citep[e.g.,][]{1991MNRAS.248P...8E,2003ApJ...596L..13B,2008A&A...481..337K,2009A&A...499...87K,2008MNRAS.389.1405K,2009ApJ...694..789T}. It is also theorized that ram pressure stripping can lead to the removal of angular momentum of gas clouds in the galaxy, funnelling the gas towards the galactic centre. This can lead to enhanced AGN activity, often observed in ram-pressure stripped galaxies  \citep[e.g.,][]{2021IAUS..359..108P,2017Natur.548..304P}.

The inner cluster regions are densely packed with galaxies, and while high velocity dispersion inhibits mergers, galaxies can still gravitationally interact during frequent flybys. This process called  \textit{harassment}, can kinematically heat the galaxies, transforming disc galaxies into ellipticals, and can transport the gas to the centre of cluster galaxies, triggering central starbursts and AGN activity \citep{1996Natur.379..613M,1998ApJ...495..139M, 1990ApJ...350...89B}. Tidal forces exerted on the galaxy during a close encounter can also remove dark matter and stars from the galaxy via \textit{tidal stripping} (e.g., \citealt{1983ApJ...264...24M,2006MNRAS.366..429R,2010MNRAS.406..729S,2012MNRAS.427.1024I,2016ApJ...833..109S,2023MNRAS.tmp.3185M}).

In contrast to overall AGN suppression inside the cluster environment, some studies suggest that the \textit{cluster outskirts} are a much more suitable environment for AGN activity. An excess of AGNs on the cluster outskirts is reported for example in \citet{2019A&A...623L..10K, 2012ApJ...754...97H, 2012AdAst2012E..32F, 2005ApJ...623L..81R}. The galaxies on the outskirts more likely belong to the in-falling population that still possesses enough cold gas to fuel star formation and AGN activity. Furthermore, the velocity dispersion of galaxies on the cluster outskirts is lower compared to the inner regions, increasing the likelihood of mergers, which are thought to be an important (although not necessarily dominant) mechanism for AGN triggering \citep[e.g.,][]{2018MNRAS.481..341S, 2008ApJS..175..356H, 2005MNRAS.361..776S,1995ApJ...448...41H, 1988ApJ...325...74S}.

To investigate the interplay between the environment and AGN activity, many studies resort to X-ray AGN selection. To get a complete sample of AGNs, not contaminated by other sources of X-ray emission, high luminosity thresholds ($\sim 10^{42}$ erg/s) are usually used \citep[e.g.,][]{2019A&A...623L..10K, 2018A&A...620A..20K, 2014MNRAS.437.1942E, 2010ApJ...723.1447H}. The abundance of such X-ray bright AGNs is low; however, modern X-ray telescopes (e.g., Chandra X-ray observatory) and their deep extra-galactic surveys opened the door for statistically meaningful analysis of the environmental dependence of AGN activity
 (for an overview of X-ray surveys, see \citealt{2005ARA&A..43..827B}). In this work, we investigate the distribution of AGNs and the relationship between AGN activity and the environment by means of cosmological hydrodynamical simulations. For a straightforward comparison with observations, we also mimic the X-ray AGN selection.

The paper is organized as follows. In Sect. \ref{simulations} we describe the {\it Magneticum} simulations. In Sect. \ref{methodology} we define clusters, used in this work, and different populations of galaxies based on their stellar mass and star formation rate. We define the distance to the nearest neighbour as a proxy for the local density and isolated and overlapping regions. We describe the calculation of AGN luminosities and define the star-forming ratio and AGN fraction, the principal quantities studied in this work. Section \ref{results} is divided into several subsections. In Sect. \ref{influence_of_neighbours}, we investigate the effects of the neighbouring clusters in a cosmological environment on the radial profiles of galaxy properties in galaxy clusters. in Sect. \ref{star_formation} we look into the radial dependence of star formation rate and the star-forming ratio and explore the effects of substructures. In Sect. \ref{sfr_local_density} we investigate the effects of local density on star formation. In Sect. \ref{AGNactivity} we discuss properties of AGN population in our simulations, focusing on the stellar mass dependence. We also investigate the radial dependence of the AGN fraction in the inner cluster regions (Sect. \ref{radial_dependence_of_AGN_fraction}) and up to large radii (Sect. \ref{AGN_fraction_at_large_r}). Section \ref{agn_and_local_density} is dedicated to the impact of the local density on AGN fraction. In Sect. \ref{comparisson_with_xray} we compare our findings with X-ray observational studies, and we conclude our paper by summarizing the main findings in Sect. \ref{conclusion}.

\section{Simulations}
\label{simulations}
For the purpose of this work we used the box2b from the \textit{Magneticum Pathfinder} suite of cosmological hydrodynamical simulations\footnote{Technical information about the \textit{Magneticum} project is available at \url{www.magneticum.org}.}, which covers a large cosmological volume with a sufficiently high number of galaxy clusters and groups for statistically meaningful analysis and has a large enough resolution to describe the behaviour inside individual galaxies realistically.

The {\it Magneticum} simulations were performed with the Tree/SPH code \textsc{Gadget-3}, a refined version of \textsc{Gadget-2} \citep{2005MNRAS.364.1105S, 2005Natur.435..629S}. It uses an improved Smoothed Particle Hydrodynamics (SPH) solver for the gas evolution (\citealt{2016MNRAS.455.2110B}) and includes the treatment of viscosity following \cite{2005MNRAS.364..753D}. The simulations take into account various physical processes such as gas cooling (\citealt{2009MNRAS.393...99W}), star formation and stellar feedback \citep[][]{2005MNRAS.361..776S, 2003MNRAS.339..289S}, UV/X-ray background heating (after \citealt{2001cghr.confE..64H}), chemical enrichment \citep{2007MNRAS.382.1050T, 2004MNRAS.349L..19T} and thermal conduction (\citealt{2014arXiv1412.6533A}). The simulations follow the SMBH evolution powering AGN feedback, based on the models from \cite{2005MNRAS.361..776S} and \cite{2005Natur.433..604D} with further improvements by \cite{2010MNRAS.401.1670F} and \cite{2014MNRAS.442.2304H}. For a concise description of the simulations, refer to \cite{2016MNRAS.463.1797D}. The {\it Magneticum} simulations assume the WMAP cosmology (\citealt{2009ApJS..180..330K}) with Hubble constant $H_0 = 70.4$ $\mathrm{km s^{-1} Mpc^{-1}}$, total matter density parameter $\Omega=0.272$, baryonic fraction of 16.8 $\%$, index of the primordial power spectrum $n = 0.963$, and normalization of the fluctuation amplitude $\sigma_8=0.809$. The \textit{hr} resolution run box2b used in this work spans a volume of (640 comoving Mpc/$h$)$^3$ and contains $2\cdot2880^3$ particles. The mass resolution is $6.9 \cdot 10^8 M_\odot/h$ and $1.4\cdot10^8M_\odot/h$ for dark matter and gas particles, respectively. The gas particles can form stellar particles with $\sim 1/4$ of their mass. Softening lengths are $3.75 \mathrm{kpc}/h$ for dark matter and gas particles and $2 \mathrm{kpc}/h$ for stellar particles. The analysis of this work relies heavily on the substructure identification, which is performed by the subhalo identifier \textsc{Subfind} \citep{2009MNRAS.399..497D, 2001MNRAS.328..726S}.

\section{Definitions}
\label{methodology}
In this work, we analyse two redshift snapshots of box2b; we will refer to $z=0.25$ as \textit{low redshift} and $z=0.90$ as \textit{high redshift}. At low redshift, \textsc{Subfind} identifies $\sim 56000$ friends-of-friends (FoF) groups with $M_{500}>10^{13}M_\odot$\footnote{The cluster size in this work is defined with the overdensity radius $R_{500}$ which is the radius where the enclosed mean mass density is $500$ times larger than the critical density of the Universe at the redshift of the cluster. Cluster mass $M_{500}$ is defined as the total mass enclosed within radius $R_{500}$.} (henceforth referred to as \textit{(all) galaxy clusters}) and 34000 at high redshift. If we restrict to massive systems with $M_{500}>10^{14}M_\odot$ (referred to as \textit{massive clusters}), we find 2000 groups at low redshift and 500 groups at high redshift. Note that there are also a few very massive clusters with $M_{500}>10^{15}M_\odot$, but they will not be considered separately in this work due to the small sample size. The centre of each cluster is defined as the position of the most bound particle and the clustercentric distance $r$ is defined relative to $R_{500}$ of each cluster. The \textsc{Subfind} algorithm also allowed us to identify substructures embedded in clusters. We were thus able to compare the behaviour of galaxies in bound subgroups containing at least 2 substructure members (i.e. galaxies) with total stellar mass $\log M_* (M_\odot)>10.15$ and single galaxies that are bound only to the main cluster halo.

Galaxies are divided into different categories based on their properties returned by \textsc{Subfind}. In terms of stellar mass $M_*$, the galaxies are split into two classes. We refer to galaxies with $\log M_* (M_\odot)> 11$ as \textit{massive} galaxies and galaxies with $10.15 < \log{M_*( M_\odot)} < 11$ as \textit{low-mass} galaxies. The lower selection limit $\log{M_*( M_\odot)}=10.15$ is approximately equal to the stellar mass at which the black hole particles are seeded in galaxies. In terms of star formation rate (SFR), the galaxies are divided into \textit{star-forming} galaxies with $\mathrm{SFR}>0.02 M_\odot/\mathrm{yr}$, and \textit{quiescent} galaxies with $\mathrm{SFR}<0.02 M_\odot/\mathrm{yr}$. Below this threshold, galaxies in our simulations have essentially negligible SFR. Note that this definition differs from the conventional criterion based on specific star formation rate (sSFR) (e.g., \citealt{2008ApJ...688..770F}), however, it ensures that both categories are sufficiently populated at all mass ranges, allowing for a meaningful comparison. While the difference in classification is negligible in low-mass galaxies, the reader should keep in mind that massive star-forming galaxies in this work may be considered quiescent according to criteria that divide galaxies based on sSFR (e.g., \citealt{2008ApJ...688..770F}).

The brightest cluster galaxies (BCGs) were identified as the closest galaxies to the cluster centre, lying at the bottom of the cluster potential. As their properties differ significantly from other cluster members we in most cases excluded them from the innermost radial bins (refer to figure captions). In some cases, we also excluded BCGs of the neighbouring clusters (see fig. \ref{Asfrquiesratio_overlapping_onecolumn} and \ref{AGNprofile_overlapping_column}).

To investigate the effects of the local environment, we define the distance to the nearest neighbour $d_n$ as the distance of a galaxy to its nearest neighbouring galaxy with $\log M_* (M_\odot)>10.15$. In large samples, this quantity correlates with the local number density of galaxies, while also enabling the identification of galaxies in the immediate proximity of another galaxy. Distance $d_n$ is given as a proper distance in kpc. 

To study the effects of the global cluster environment, we count the number of AGNs and galaxies in spherical shells at different clustercentric distances $r$ (in $R_{500}$). To increase the sample size, the radial bins are stacked together and galaxies are counted in all clusters, similarly to \cite{2019A&A...623L..10K}, \cite{2018A&A...620A..20K} and \cite{2014MNRAS.437.1942E}. Note that some galaxies are close to several clusters and may be considered multiple times. The confidence intervals shown in this work (e.g., see Fig. \ref{Asfrquiesratio_overlapping_onecolumn}) refer to the statistical error of the stacked profiles. In cases where the \textit{star-forming ratio} between the number of star-forming galaxies $N_{\rm sfr}$ and the number of quiescent galaxies $N_{\rm quies}$ is shown, the confidence interval is obtained by assuming that $N_{\rm sfr}$ and $N_{\rm quies}$ are independent and are drawn from Poisson distribution. For an in-depth discussion about the ratio of Poisson variables refer to \cite{2006ApJ...652..610P}. In cases where the fraction of galaxies containing an AGN (henceforth \textit{AGN fraction}) is plotted, the errors are obtained by assuming that the number of AGNs is drawn from the binomial distribution with the known total number of galaxies.

\begin{figure*}
\centering
\includegraphics[width=\textwidth]{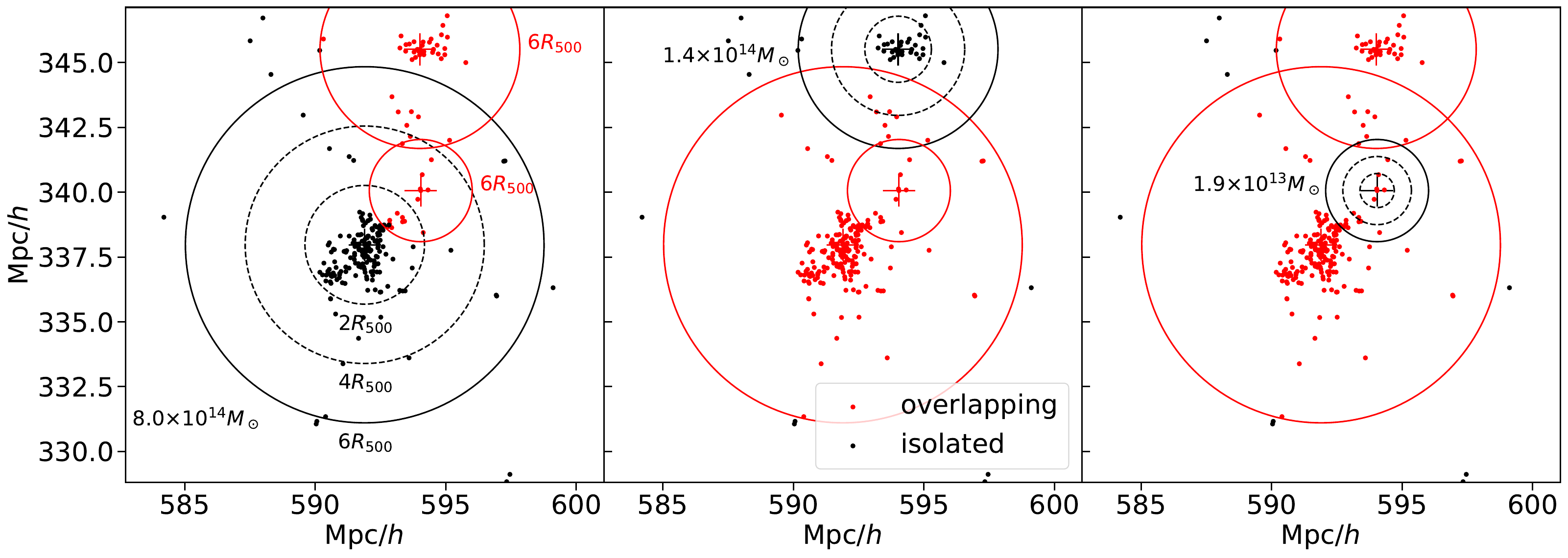}
\caption{The three panels depict the region of 18 x 18 x 2 comoving Mpc/$h$ at redshift 0.25 containing 3 clusters. Dots represent comoving positions of galaxies. Different panels depict the definition of the overlapping (red) and the isolated (black) regions in each cluster, marked with black circles that represent its $2 R_{500},4 R_{500}$ and $6 R_{500}$. Red circles represent $r_{ov}=6 R_{500}$ of the neighbouring clusters, and the crosses mark the cluster centres. Mass $M_{500}$ of the considered cluster is displayed in each panel.}
\label{overlaps_definition}
\end{figure*}

Since galaxy clusters are not isolated but are placed in a large-scale cosmological environment it is insightful to study how the radial profiles of their properties (i.e. AGN fraction \& star-forming ratio) are influenced by their neighbouring clusters. Hence we define the \textit{overlapping} and \textit{isolated} population of galaxies. The overlapping population comprises the galaxies that are closer than the overlapping radius $r_{ov}$ to some neighbouring cluster (with $M_{500}>10^{13} M_\odot$). All the other galaxies are assigned to the isolated population. In this work, we set $r_{ov}=6 R_{500}$, where $R_{500}$ refers to each neighbouring cluster. The rather large choice of $r_{ov}$ ensures that the isolated regions are not influenced by any nearby cluster, while it still divides the galaxies into two sufficiently populated samples for a statistically meaningful comparison. In Fig. \ref{overlaps_definition}, we demonstrate the definition of the overlapping and isolated population in three clusters. Note that some galaxies may simultaneously belong to the overlapping population of one cluster and the isolated population of another cluster.

To determine whether a galaxy in the simulation hosts an AGN we first match the black holes with their host galaxies by their positions. The bolometric X-ray luminosity $L_{bol}$ of the AGN is then calculated as $L_{bol}=\epsilon_r \dot{M} c^2$ where the $\dot{M}$ represents the black hole mass accretion rate, $\epsilon_r$ the efficiency which is set to 0.1 (see \citealt{2013ApJ...767...37M} and references therein) and $c$ the speed of light. Bolometric luminosity is then converted to X-ray luminosity by applying the bolometric corrections proposed by \cite{2004MNRAS.351..169M}. Any galaxy hosting a black hole brighter than $10^{42}$ erg/s in 0.5-10 keV band is considered to host an (X-ray bright) AGN. This simple luminosity estimate allows for a straightforward comparison to observational studies (e.g., \citealt{2018A&A...620A..20K}) while keeping the intuitive connection with the mass accretion rate $\dot{M}$ and thus directly reflecting the environmental effects.

\section{Results}
\label{results}
\subsection{Influence of the large-scale environment}
\label{influence_of_neighbours}
First, we will investigate how the radial dependence of galaxy properties in clusters is influenced by the neighbouring clusters in a cosmological environment. To this scope, let's assume that some property of cluster galaxies $p(r)$ (e.g., the AGN fraction or the star-forming ratio of galaxies) only depends on the distance $r$ to its "parent" cluster and that the profiles $p(r)$ of all clusters are identical if re-scaled to $R_{500}$ of each cluster. For this example, we further assume that when several clusters are brought together to overlap in a cosmological environment, their profiles remain the same - each galaxy still belongs to its parent cluster. The average property $\langle p \rangle$ of galaxies in volume $\Delta V$ at position $i$ is calculated by summing the contributions of the central and other surrounding clusters. Each cluster $j$ at a distance $r_{ij}$ (relative to $R_{500}$ of $j$-th cluster) from position $i$ contributes the number of galaxies with properties $p(r_{ij})$, proportional to the number density $\mathrm{d}N/\mathrm{d}V (r_{ij})$. In other words,  $\langle p \rangle$ at position $i$ is calculated as the average of the profiles  $p(r_{ij})$ of all clusters, weighted by their number density at position $i$. In this work, we will not be interested in the average $\langle p \rangle$ at a single point $i$, but in the average $\langle p \rangle$ in the \textit{radial shells} at a distance $r$ from the centre of the cluster. This can be achieved by sampling the volume of the spherical shells at a distance $r$ from the cluster centres with random points $i$. The average property of galaxies $\langle p \rangle$ at $r$ can be calculated as follows:

\begin{equation}
\label{eq:weightedaverage}
    \langle p \rangle=\frac{\sum\limits_{i} \sum\limits_{j} p(r_{ij}) \frac{\mathrm{d}N}{\mathrm{d}V} (r_{ij})}
{ \sum\limits_{i} \sum\limits_{j} \frac{\mathrm{d}N}{\mathrm{d}V} (r_{ij})}.
\end{equation}

The sum in equation \eqref{eq:weightedaverage} runs over all points $i$ in the radial shells at some distance $r$ from the central clusters, and at each point, over the contributions of all clusters $j$. If the radial dependence of the property $p$ and number density of galaxies in an isolated galaxy cluster are known, equation \eqref{eq:weightedaverage} can be used to estimate how the profile would appear in the large-scale environment when influenced by the identical profiles of other clusters. While equation \eqref{eq:weightedaverage} can be used for the profile of any property of galaxies $p$, it is already instructive to define $p(r)=r$. The average $\langle p \rangle = \langle r \rangle$, obtained with equation \eqref{eq:weightedaverage}, can be interpreted as a characteristic clustercentric distance of galaxies at a distance $r$. It tells us the expected distance of galaxies at a distance $r$ from the \textit{central} cluster to \textit{any} (central or other) cluster. If a cluster is in complete isolation, that means that $\langle r \rangle=r$; all galaxies belong to the central cluster and have properties characteristic of the distance $r$. However, if a cluster is surrounded by other clusters, the  $\langle r \rangle$ decreases at large $r$; some galaxies may belong to the inner regions of other clusters in the vicinity, with properties characteristic of low $r$. We emphasize that $r$ refers to the distance relative to the $R_{500}$ of each parent cluster (central and other) and that the characteristic distance $\langle r \rangle$ should be interpreted merely as a proxy for the properties of galaxies, rather than the measure of the local density.

To estimate the characteristic distance of galaxies $\langle r \rangle$ as a function of $r$, we used the following procedure. In each cluster, we generated a set of 80000 uniformly distributed random points in the sphere of radius $8 R_{500}$\footnote{We generated the same number of random points in each cluster, regardless of its size. To account for different number densities of generated points, we thus weighted each point $i$ with the volume of its central cluster ($\propto R_{500}^3$).}. We then selected the points in a narrow radial shell at a distance $r$ from the centre of the considered (central) cluster. In each of those points $i$, we added the contributions of all clusters $j$, based on their distance from the point $i$ and their number density at this distance (see equation \ref{eq:weightedaverage}). For the number density $\mathrm{d}N/\mathrm{d}V (r_{ij})$, we assumed the Navarro-Frenk-White (NFW) profile \citep{1997ApJ...490..493N, 1996ApJ...462..563N, 1995MNRAS.275..720N}. The parameters of the profile were obtained by fitting the profile to the number density of galaxies with  $\log M_* (M_\odot)> 10.15$ (mostly low-mass galaxies) in the simulation box. NFW profile was assumed between $0.15 R_{500}$ and $10 R_{500}$. Below $0.15 R_{500}$, the number density was set to a constant value to avoid the unrealistically high contribution of cluster centres, where the NFW profile diverges. Above $10R_{500}$, the number density was set to 0 to neglect the influence of individual clusters over large cosmological distances. With this assumption, the sum over \textit{all} clusters effectively reduces to a sum over all clusters comprised within $10R_{500}$ from the central one, which we call hereafter \textit{neighbouring} clusters. We verified that our conclusions do not depend on the choice of those two threshold distances. After summing the contributions of all central and neighbouring clusters in each point, we summed the contributions of all points $i$ at a distance $r$ from their central clusters to obtain average $\langle r \rangle$ at a distance $r$ (see equation \ref{eq:weightedaverage}).

\begin{figure}
\centering
\includegraphics[width=\textwidth/2]{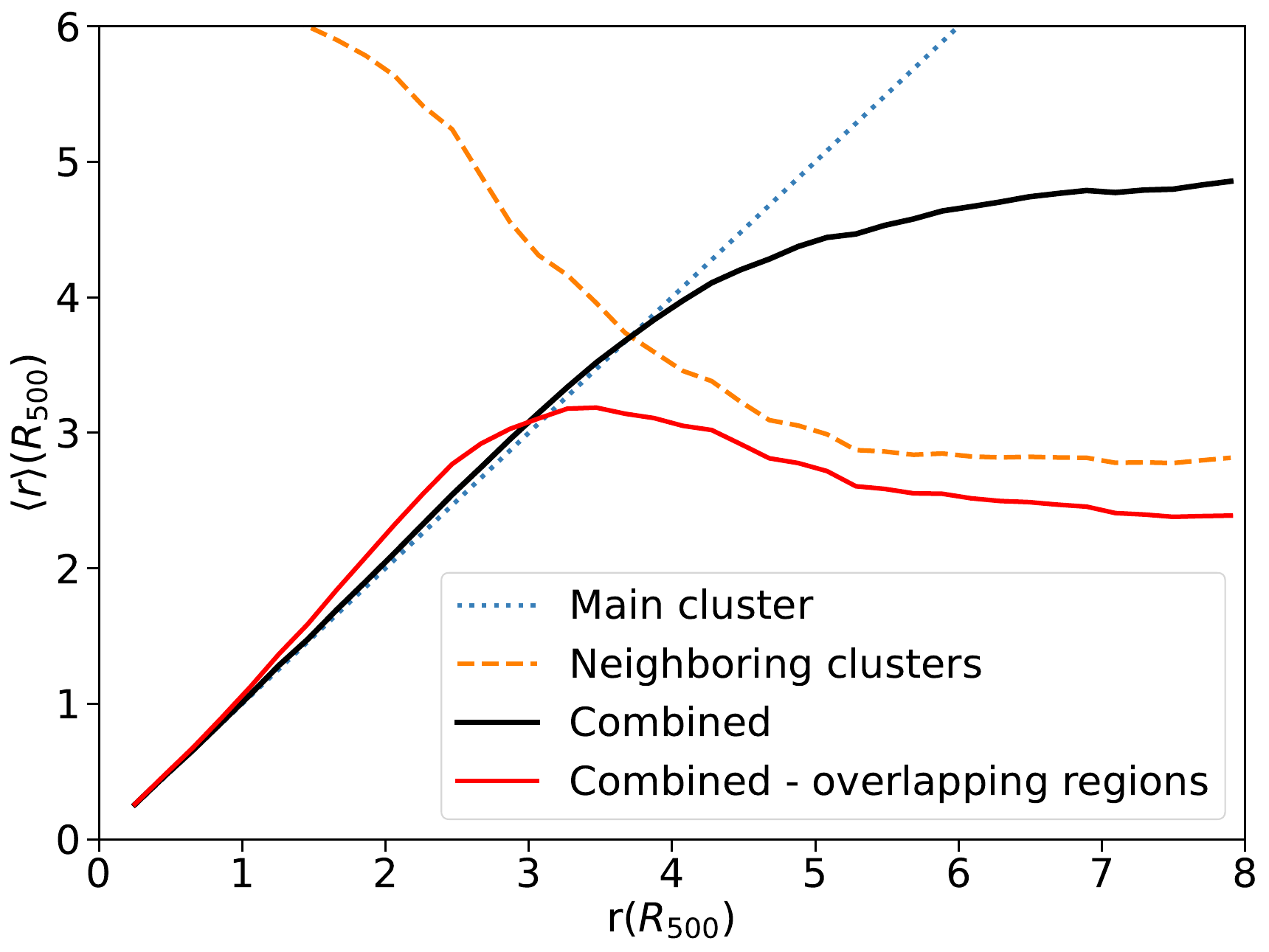}
  \caption{Characteristic distance $\langle r \rangle$ (equation \eqref{eq:weightedaverage} where $p(r)=r$) as a function of the clustercentric distance computed at low redshift ($z=0.25$) with procedure and definition of neighbouring clusters described in section \ref{influence_of_neighbours}. Clustercentric distance $r$ is given in $R_{500}$ of the central clusters and the characteristic distance $\langle r \rangle$ is given in $R_{500}$ of all clusters in the calculation (central or neighbouring). In the calculation of the blue dotted profile, only the contributions of the central clusters are considered, and all distances contributing to $\langle r \rangle$ are given relative to $R_{500}$ of the central cluster. In the calculations of the dashed orange profile, the contributions of neighbouring clusters are considered - each neighbouring cluster contributes a number of galaxies with distances, given relative to its $R_{500}$. The black profile is the combined profile of central and neighbouring clusters. As the distances are re-scaled to different $R_{500}$, $\langle r \rangle$ should be seen as a proxy for the properties of galaxies at some distance $r$, rather than the measure of the local density. Red solid line is calculated in the same way as the combined profile (black solid line) but in the overlapping regions only.}

     \label{overlaps_MC}
\end{figure}

Characteristic distance $\langle r \rangle$ as a function of $r$, obtained with the described procedure, is plotted in Fig. \ref{overlaps_MC} (black solid line). We also repeated the computation by neglecting the contribution of the central clusters (orange dashed line). The blue dotted line represents the case where only the contribution of the central clusters is considered ($\langle r \rangle=r$). For the red solid line, we have not considered all generated points $i$ but only those in the overlapping regions (see section \ref{methodology}). 
In the combined profiles of all (central and neighbouring) clusters, we can see how in the inner regions ($\lesssim 4 R_{500}$) the contribution of neighbouring clusters causes a relative increase of $\langle r \rangle$. This is unsurprising as at those radii, the average distance to the neighbouring clusters ($r_{ij}$) is larger than the distance to the central cluster $r$. As the number density of the central cluster decreases at large radii ($\gtrsim 4 R_{500}$), the contributions of the inner regions of neighbouring clusters become increasingly important, causing a relative decrease of $\langle r \rangle$. This is evident in the combined profile of all regions (black line) and the profile of overlapping regions (red line). In the latter, the transition from the central cluster dominating the inner regions to the significant contribution of neighbours at large $r$ appears as a peak; galaxies between 3 and 4 $R_{500}$ have, on average, the highest $\langle r \rangle$, i.e., are the furthest away from any cluster in terms of $R_{500}$. The exact position of the peak depends on the definition of the overlapping regions (overlapping radius $r_{ov}$). This peak implies that if some property $p$ has a distinctively lower value in the inner cluster regions, this may appear as a peak on the cluster outskirts simply because the galaxies there are the furthest away from any cluster regions. This peak is a consequence of geometrically overlapping cluster regions and does not necessarily imply any physical excess on the cluster outskirts with respect to the field. Figure \ref{overlaps_MC} shows the behaviour at low redshift ($z=0.25$). We verified that the behaviour at high redshift ($z=0.90$) is qualitatively similar, despite the peak of the overlapping regions is found at slightly lower $r$, at around $3 R_{500}$.

\subsection{Star formation and the global environment}
\label{star_formation}
\begin{figure}
\centering
\includegraphics[width=\textwidth*19/40]{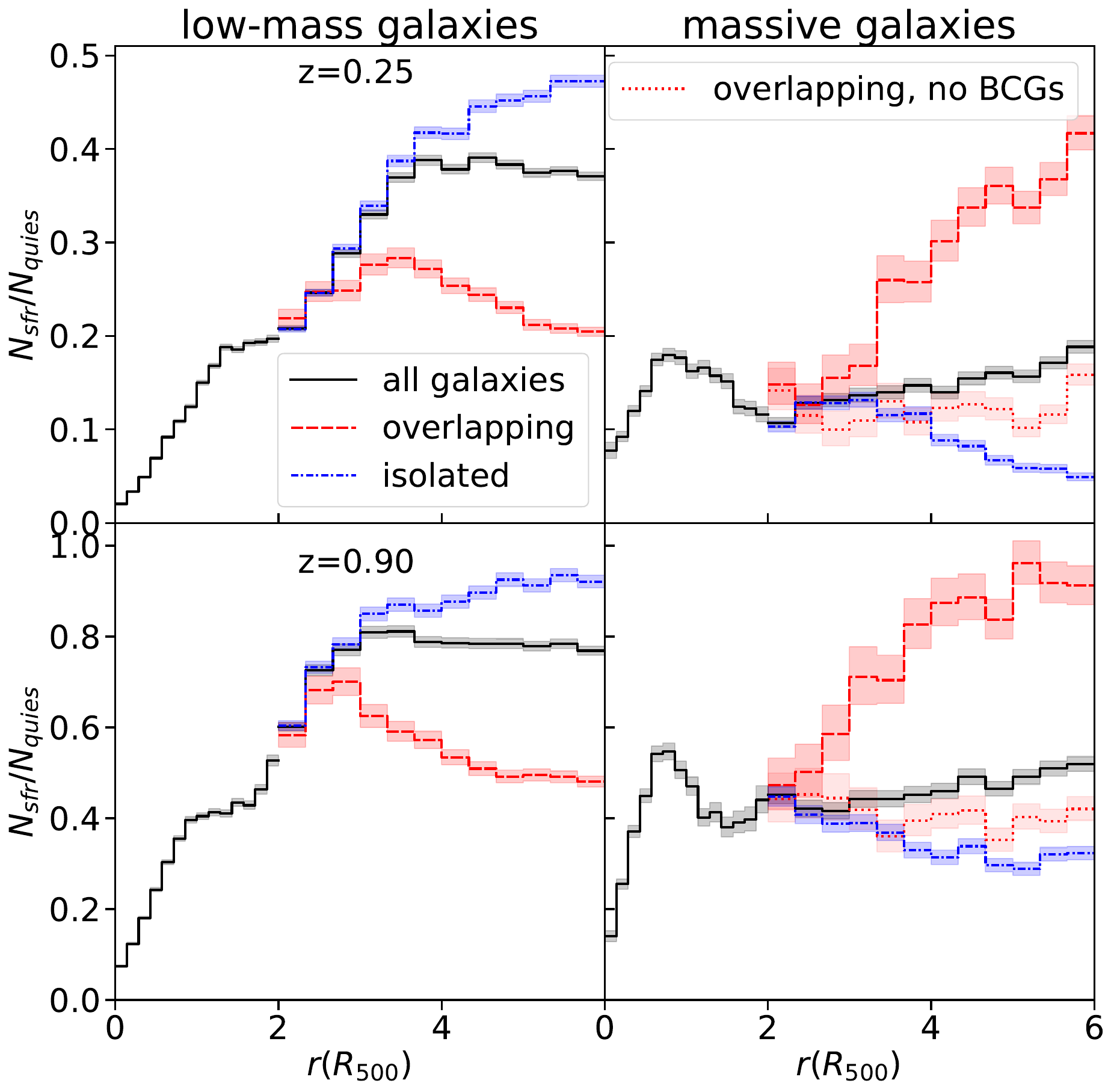}
  \caption{Ratio between the number of star-forming ($\mathrm{SFR}>0.02 M_\odot/\mathrm{yr}$) and quiescent galaxies as a function of clustercentric distance $r$. Low-mass galaxies ($10.15 < \log M_* (M_\odot)<11$) are shown on the left and massive galaxies ($\log M_*(M_\odot)>11$) on the right. Results are shown at low redshift (upper panels) and high redshift (lower panels). For $r>2R_{500}$ we separately plot the profiles of the overlapping and isolated regions ($r_{ov}=6 R_{500}$, see Sect. \ref{methodology}). On the right panels, the dotted line represents the massive overlapping populations without the BCGs of the neighbouring clusters, central BCGs are excluded in all plots. The coloured bands represent the 68 \% confidence intervals obtained with Poisson statistics (see Sect. \ref{methodology}). }
\label{Asfrquiesratio_overlapping_onecolumn}
\end{figure}

To get some insight into processes governing the star formation rate, we plot the star-forming ratio as a function of clustercentric distance $r$ (Fig. \ref{Asfrquiesratio_overlapping_onecolumn}). The ratio is plotted at high and low redshift and for low-mass and massive galaxies divided into overlapping and isolated populations as described in Sect. \ref{methodology}. In low-mass galaxies (left panels), we can notice a strong decline of the star-forming ratio towards the cluster centre at low and high redshift, which is in agreement with observations \citep[e.g.,][]{2014ApJ...783..136C, 2012MNRAS.424..232W}. The star-forming ratio of massive galaxies (right panels, isolated regions) behaves considerably differently - it increases towards the inner cluster regions. The overall increase features two peaks, one located on the outskirts (around 3 $R_{500}$ at low $z$) and one in the inner cluster regions, between 0.5 and 1 $R_{500}$, which is followed by a rapid decline towards the centre. The drastic difference in the behaviour between low-mass and massive galaxies is related to the role of mass quenching in massive galaxies. Whereas low-mass galaxies mostly succumb to environmental quenching mechanisms and would likely still be forming stars outside the cluster, massive galaxies are already quenched prior to their infall. Infalling massive galaxies then have star formation briefly reignited due to the pressure inside the ICM (see section \ref{introduction}).

This behaviour has been thourughly investigated in a smaller simulations box from the Magneticum suite in a dedicated study by \citealt{2019MNRAS.488.5370L}, for a smaller sample of galaxies tracked during their infall trajectories. \citealt{2019MNRAS.488.5370L} found that the specific star formation rate of galaxies with $M_* > 1.5 \cdot 10^{10}M_{\odot}$ increased briefly after the galaxies crossed the virial radius (see Figure 10 in \citealt{2019MNRAS.488.5370L}). They also report that the levels of star formation in massive galaxies are lower than in low-mass galaxies prior to infall and gradually decline, indicating the role of mass quenching. Low-mass galaxies on the other hand are quenched rapidly after crossing the virial radius. 

In Fig. \ref{Asfrquiesratio_overlapping_onecolumn}, we can also see how the behaviour in the inner cluster regions is reflected in the profile of the overlapping population and, consequently, in the combined profile. The central decline of the star-forming ratio found in low-mass galaxies (left panels) is reflected in the decrease at large $r$ ($r>4R_{500}$), where inner regions of neighbouring clusters are considered, as was demonstrated in Sect. \ref{influence_of_neighbours}. The peak of the star-forming ratio in low-mass galaxies in the overlapping regions around $3 R_{500}$ is analagous to that predicted in Fig. \ref{overlaps_MC} and can thus be explained by geometric overlaps of the cluster regions and superposition of their profiles; it does not imply the excess of star formation on the outskirts of each cluster. Analogously, an increase of the star-forming ratio in massive galaxies (right panels) in the overlapping regions at large $r$ ($r>4R_{500}$) results from the central increase in the neighbouring clusters. Note that the star-forming ratio of the massive overlapping population is profoundly influenced by the neighbouring BCGs (compare dashed lines with BCGs and dotted lines with BCGs excluded). We also verified that despite the small sample size, signs of the secondary peak at $\sim 3 R_{500}$ in Fig. \ref{Asfrquiesratio_overlapping_onecolumn} remain even if a sub-sample of isolated clusters (as opposed to the isolated regions of any cluster) and if only single galaxies (not substructure members) are selected.

\begin{figure}
\centering
\includegraphics[width=\textwidth*19/40]{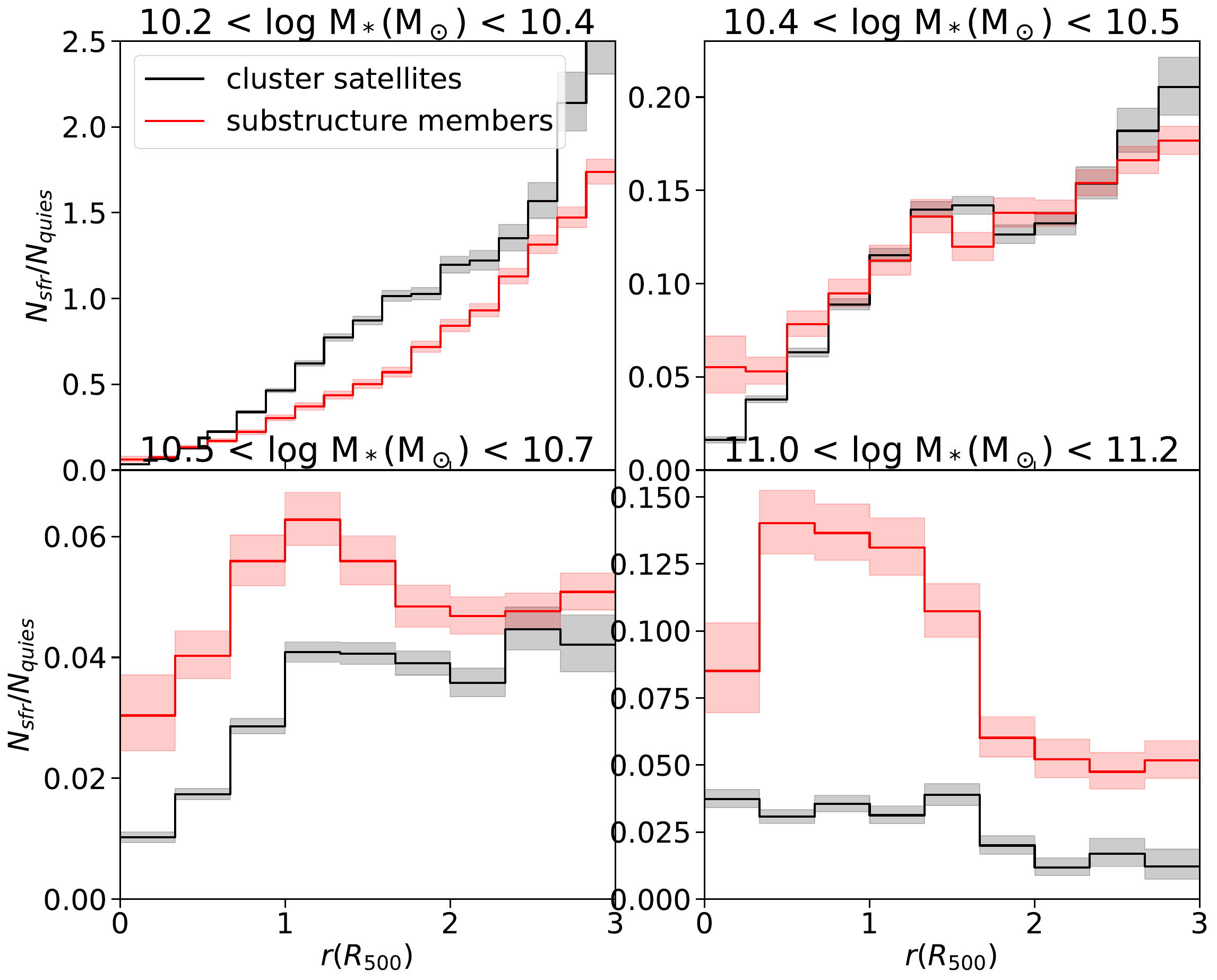}
  \caption{Star-forming ratio as a function of $r$ for galaxies in 4 different $M_*$ ranges (different panels). The red line represents galaxies that are embedded in substructures with at least two galaxies with $\log M_* (M_\odot)>10.15$. The black line represents single galaxies that orbit the cluster alone.
  Central BCGs are excluded. The coloured bands represent the 68 \% confidence intervals obtained with Poisson statistics (see Sect. \ref{methodology}). }
\label{sfr_substructures_panels}
\end{figure}

The star-forming ratio and its increase in the inner regions depends on the substructure membership. In Fig. \ref{sfr_substructures_panels}, we plot the star-forming ratio in the inner cluster regions as a function of clustercentric distance for the substructure members (red line) and single galaxies that belong directly to the main cluster (black line). Galaxies in narrow $M_*$ ranges are selected in each panel to mitigate the effects of mass segregation. The star-forming ratio in massive substructure members (bottom panels) is generally higher and shows a much more pronounced peak around $R_{500}$ than in their isolated counterparts. This is most likely due to a combination of two effects. The group atmosphere can shield the galaxy from the quenching effects of the intracluster medium. \citet{2019MNRAS.488.5370L} found that massive central galaxies of in-falling groups remain star-forming the longest after the in-fall. They attribute this to their higher stellar mass and the group environment mitigating the effects of ram-pressure stripping. The other effect contributing to the difference between substructure members and cluster satellites in Fig. \ref{sfr_substructures_panels} is orbital selection. Since galaxy groups inside the cluster environment usually get unbound soon after the pericentric passage, they are likely in-falling into the cluster \citep[e.g.,][]{2023MNRAS.518.1316H,2019MNRAS.490.3654C}. By examining orientations of velocities of galaxies with $0<r<3R_{500}$ (in all mass ranges used in Fig. \ref{sfr_substructures_panels}), we checked that a larger fraction of substructure members have radially inward-oriented velocities with dominant radial components, indicating a larger fraction of the in-falling population than in the single galaxy sample. Quantifying how much both of those effects contribute to the shape of the radial profiles shown in Fig. \ref{sfr_substructures_panels} would, however, require a more elaborate analysis, including tracking individual galaxies during their infall, and is beyond the scope of this work.  The overall increase of the star-forming ratio in clusters with rich substructure is also found in observations (e.g., \citealt{2014ApJ...783..136C}).

Low-mass substructure members (top panels), on the other hand, exhibit a lower star-forming ratio than their isolated counterparts. This is consistent with environmental quenching effects in denser environments (see also Sect. \ref{sfr_local_density}). It was shown by \citealt{2019MNRAS.488.5370L} that low-mass cluster satellites that are not substructure members have the highest level of star formation prior to in-fall.

\begin{figure*}
\centering
\includegraphics[width=\textwidth*9/11]{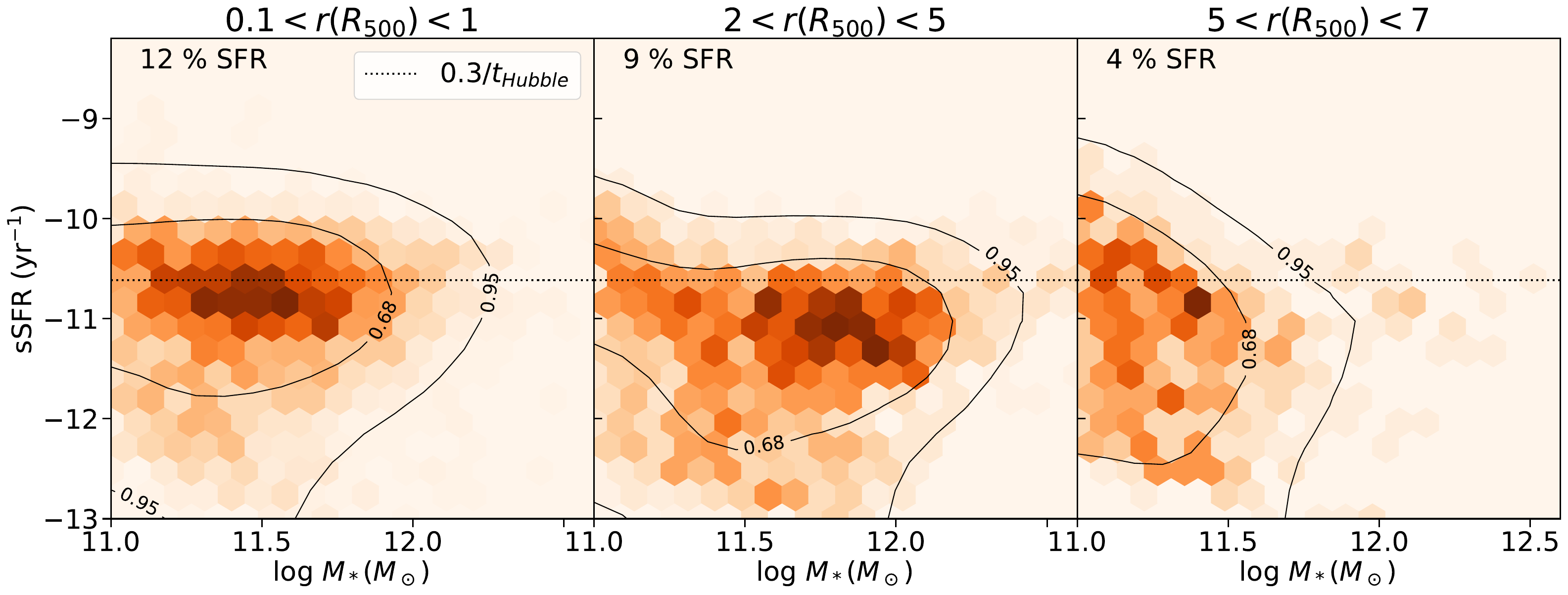}
\caption{Distribution of stellar mass $M_*$ and specific star formation rate sSFR of massive galaxies at different clustercentric distances $r$. Only isolated regions were selected. The dashed line indicates the conventional division between star-forming and quiescent used in some other works (\citealt{2008ApJ...688..770F}). In the top left corner of each panel, the fraction of star-forming galaxies ($\mathrm{SFR}>0.02 M_\odot/\mathrm{yr}$) in each region is shown. The contours show regions containing $68\%$ and $95\%$ of galaxies. The color range is normalized to the total number of galaxies in each panel. Results are shown at low redshift ($z=0.25$).}
\label{mst_ssfr}
\end{figure*}

Aside from the star-forming ratio it is also instructive to see how the sSFR value of star-forming galaxies varies with $r$. In Fig. \ref{mst_ssfr} the $M_*-\mathrm{sSFR}$ distribution of massive galaxies is shown in different radial ranges. The expected sSFR value depends on the environment. Star-forming galaxies of some mass $M_*$ have higher expected sSFR in the inner cluster regions (below $R_{500}$, left panel) than on the outskirts (above $2R_{500}$, middle and right panel), which is in agreement with what is found in IllustrisTNG simulations (\citealt{2019MNRAS.489..339H}). This also means that the expected star formation rate of massive galaxies roughly traces the star-forming ratio and that the galaxy mass segregation cannot be the sole reason for the observed trend. We also examined the $M_*-\mathrm{sSFR}$ distribution of low-mass galaxies (not shown explicitly) and, in contrast to massive galaxies, found no noticeable changes of sSFR with $r$, despite the drastic $r$ dependence of the star-forming ratio, shown in Fig. \ref{Asfrquiesratio_overlapping_onecolumn}. In other words, low-mass galaxies are more likely to be quiescent in the cluster environment than in the field, but if they manage to retain conditions for star formation, the sSFR tends to be on the same level, regardless of the global environment ($r$). The lack of environmental dependence of the sSFR of star-forming galaxies, despite the strong environmental dependence of the star-forming ratio is also reported in observations \citep[e.g.,][]{2010ApJ...721..193P,2012ApJ...746..188M}. In our case, it is likely a consequence of rapid quenching time scales of galaxies in our simulations (see \citealt{2019MNRAS.488.5370L}) which explains the absence of galaxies in the transition phase with low sSFR.

\subsection{Star formation and local density}
\label{sfr_local_density}

\begin{figure}
\centering
\includegraphics[width=\textwidth*19/40]{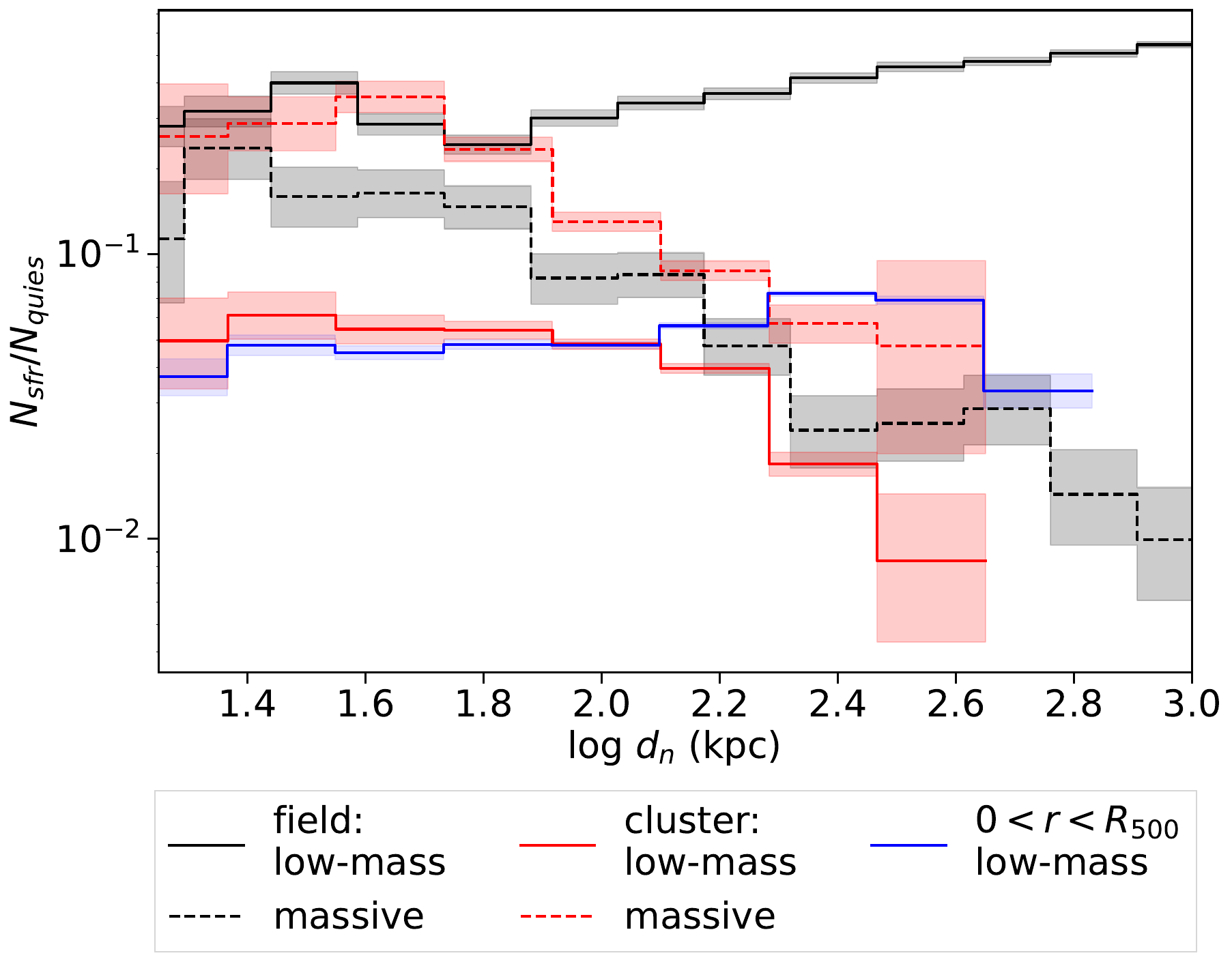}
  \caption{Ratio between the number of star-forming ($\mathrm{SFR}>0.02 M_\odot/\mathrm{yr}$) and quiescent galaxies as a function of the distance to the nearest neighbour $d_n$. Results are shown at low redshift for low-mass (solid line) and massive galaxies (dashed line). In black, we show the ratio of the field galaxies (galaxies in the isolated regions between $6 R_{500}$ and $7R_{500}$ of each cluster) and in red the ratio of galaxies in the inner cluster regions ($0.25 R_{500}<r<0.5R_{500}$). Narrow radial bins are selected to disentangle the local density effects from radial dependence; their importance is demonstrated with a blue solid plot, depicting low-mass population in a  wider radial bin ($0 <r< R_{500}$). BCGs are excluded in all plots. The coloured bands represent the 68 \% confidence intervals obtained with Poisson statistics (see Sect. \ref{methodology}).}
     \label{dnb-sfrquiesratio-narrowregions_single}
\end{figure}

We also checked how the star-forming ratio depends on the local density, traced by the distance to the nearest neighbour $d_n$. The star-forming ratio as a function of $d_n$ is shown in Fig. \ref{dnb-sfrquiesratio-narrowregions_single} for low-mass and massive galaxies at constant $r$ in the field and in the inner cluster regions. The star-forming ratio of massive galaxies increases with local density in clusters and the field, which may indicate the importance of mass quenching. In the hierarchical structure formation framework, galaxies tend to migrate from the less dense environment to denser clusters and groups and unlikely vice versa. That means a galaxy in isolation has probably formed and evolved there without being subjected to environmental quenching in the dense environments. On the contrary, a denser environment seems to either trigger star formation in quenched massive galaxies or prevent the cessation of star formation - not just in the dense cluster environment (as already discussed in section \ref{star_formation}), but wherever the local density is elevated (see dashed black line in Fig. \ref{dnb-sfrquiesratio-narrowregions_single}). We also found that the increase of star-forming ratio of massive galaxies with local density inside the cluster environment is more pronounced if they are substructure members.

While low-mass galaxies display similar behaviour in clusters, the trend is different in the field; low-mass galaxies have the highest star-forming ratio if they are far from their neighbours, which again points to the role of environmental quenching. In other words: a low-mass galaxy is more likely to form stars if it is in the field, far from any aggregations of galaxies that generally suppress star formation but if it does find itself inside the cluster environment, it is more protected from the environmental quenching if it is in a locally very dense environment. We verified that galaxies in a locally very dense environment were more likely to be substructure members.

We should also stress that since $d_n$ is highly correlated with $r$ it was crucial to select galaxies in very narrow radial bins in Fig. \ref{dnb-sfrquiesratio-narrowregions_single} to disentangle the effects of local density and the global cluster environment. This is demonstrated with the blue solid line, where a wider radial bin was chosen. The increase of star-forming ratio with local density found in narrow radial bins ($0.25 R_{500}<r<0.5R_{500}$, red solid line) is not noticeable anymore if the width of the radial bin is increased to $0 <r< R_{500}$.

\subsection{AGN population}
\label{AGNactivity}
\begin{figure}
\centering
\includegraphics[width=\textwidth*19/40]{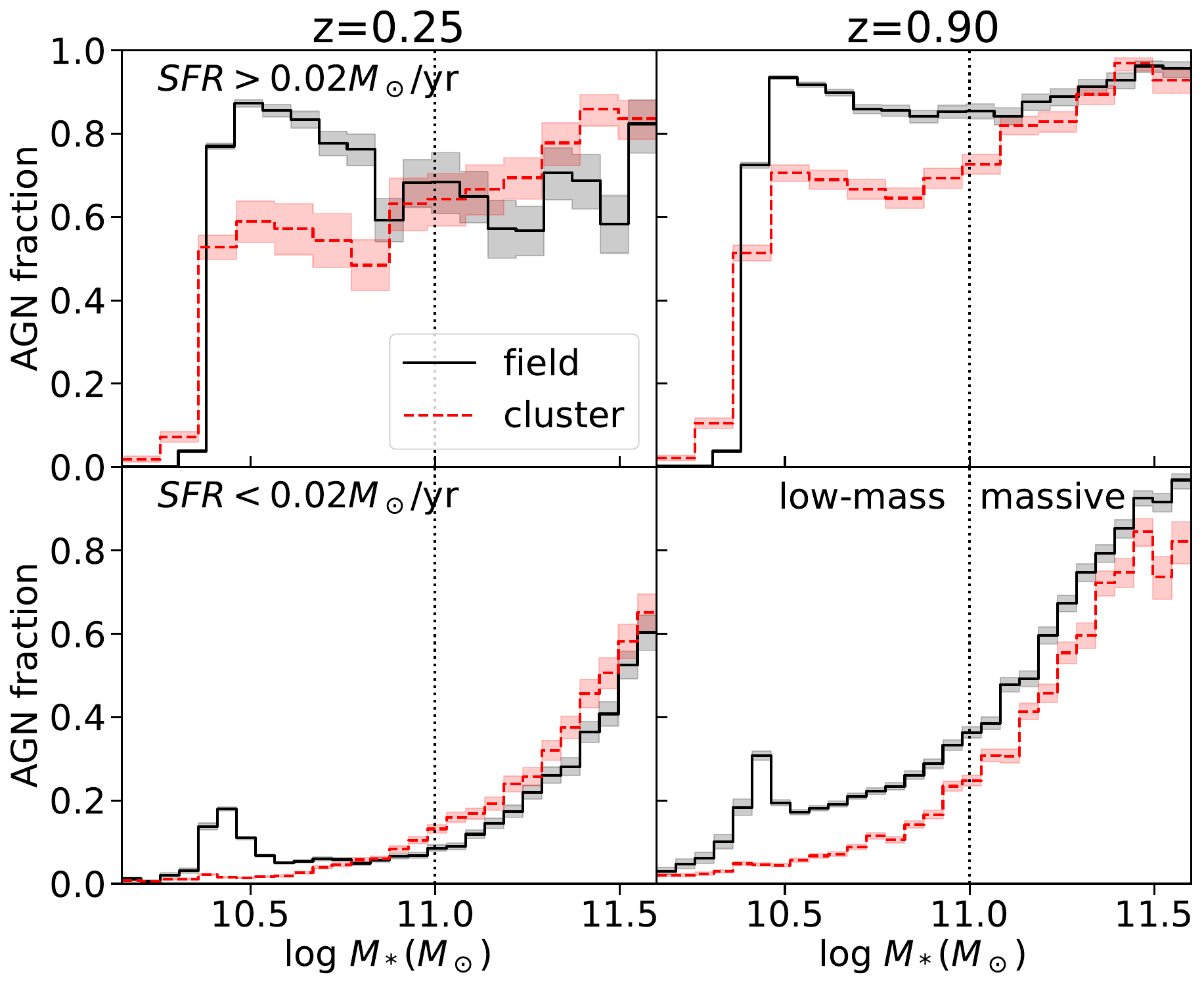}
  \caption{AGN fraction as a function of the stellar mass $M_*$ of star-forming (upper panels) and quiescent galaxies (lower panels) at high redshift (right panels) and low redshift (left panels). The solid line represents the AGN fraction of the field galaxies (galaxies in the isolated regions between $6 R_{500}$ and $7R_{500}$ of each cluster) and the dashed line the ratio of galaxies in the cluster centre (below 0.4 $R_{500}$). The dotted vertical line divides galaxies into low-mass and massive galaxies as defined in this work. The coloured bands represent the 68 \% confidence intervals obtained with binomial statistics (see Sect. \ref{methodology}). }
     \label{AGN_MSTprofile}
\end{figure}

In the following sections, we investigate the fraction of galaxies hosting an X-ray bright AGN (henceforth AGN fraction, see Sect. \ref{methodology}). Before delving into the radial dependence of the AGN fraction, we first discuss a few general properties of AGNs and their hosts in {\it Magneticum} simulations. For an in-depth discussion, refer to \cite{2018MNRAS.481.2213B}, \cite{2015MNRAS.448.1504S} and \cite{2014MNRAS.442.2304H}; here, we summarize the most important trends and point out the limitations of the simulations. In Fig. \ref{AGN_MSTprofile}, we show the AGN fraction as a function of $M_*$ in star-forming and quiescent galaxies at low and high redshift in the field and the cluster regions. First, we can see that the AGN fraction is generally higher at high redshift (right panels) than at low redshift (left panels), regardless of the environment or stellar mass. This indicates the anti-hierarchical growth of the luminous AGNs; the comoving number density of AGNs increases with time, peaks around $z=1-2$, and then declines for $z<1$ due to the decreasing cold gas content in the vicinity of massive black holes (see \citealt{2014MNRAS.442.2304H}). Moreover, we can see that at masses larger than the black hole seeding stellar mass ($\log M_* (M_\odot) \approx 10.15$) the AGN fraction rapidly rises, reaching the peak at $\log M_* (M_\odot) \approx 10.4$ then falling off before rising again at large $M_*$. The peak is not observed in nature and is a consequence of the black hole sub-grid model; black hole particles are seeded well below the $M_{\bullet}-M_{*}$ relation \citep[e.g.,][]{1998AJ....115.2285M} and thus undergo a phase of very efficient AGN feedback. They accrete below the Eddington limit and are unable to suppress the gas cooling. Thus, the black holes grow rapidly until they reach the $M_{\bullet}-M_{*}$ relation, where they abruptly quench the star formation and halt AGN activity in the host galaxy \citep[see][]{2015MNRAS.448.1504S}. Note that the mass $M_*$ corresponding to the peak in Fig. \ref{AGN_MSTprofile} is determined by our choice of black hole seeding stellar mass. Quenching due to AGN feedback also causes an accumulation of galaxies and a peak in the stellar mass function at $\log M_* (M_\odot) \approx 10.5$, not found in observations (e.g., see Figure 4 in \citealt{2014MNRAS.442.2304H}). The properties of low-mass galaxies in this work should therefore be taken with a grain of salt, although we still show them to demonstrate the effects of the cluster environment. In Fig. \ref{AGN_MSTprofile}, we can see that in contrast to field galaxies, no peak of AGN fraction at $\log M_* (M_\odot) \approx 10.4$ is observed within clusters, demonstrating a decisive role of the environment on the AGN fraction in low-mass galaxies. We can also see that the AGN activity is suppressed more gradually at $M_*$ larger than the peak value if a galaxy retains conditions for star formation (upper panels) compared to quiescent galaxies (bottom panels). In massive galaxies  ($\log M_* (M_\odot) > 11$), on the other hand, the decisive factor for AGN activity becomes stellar mass $M_*$. The relative differences in AGN fraction between massive galaxies in clusters and in the field are not as drastic as in low-mass galaxies. On the other hand, even small changes of $M_*$ can overshadow the difference in AGN fraction between the two environments. The rapid increase of AGN fraction with $M_*$ can be readily explained with the correlation between stellar mass and the black hole mass \citep[e.g.,][]{ 1998AJ....115.2285M} and with Bondi model for mass accretion rate used in {\it Magneticum} simulations in which the accretion rate scales with the square of the black hole mass \citep{2005MNRAS.361..776S, 1952MNRAS.112..195B}. Galaxies with sufficiently massive black holes can make it above the AGN selection threshold even if they are in gas-poor environment according to our simple luminosity estimate described in Sect. \ref{methodology}. Note that some of those galaxies may in reality be considered radio galaxies due to their low Eddington ratio.

\subsection{Radial dependence of the AGN fraction}
\label{radial_dependence_of_AGN_fraction}

\begin{figure}
\centering
\includegraphics[width=\textwidth*19/40]{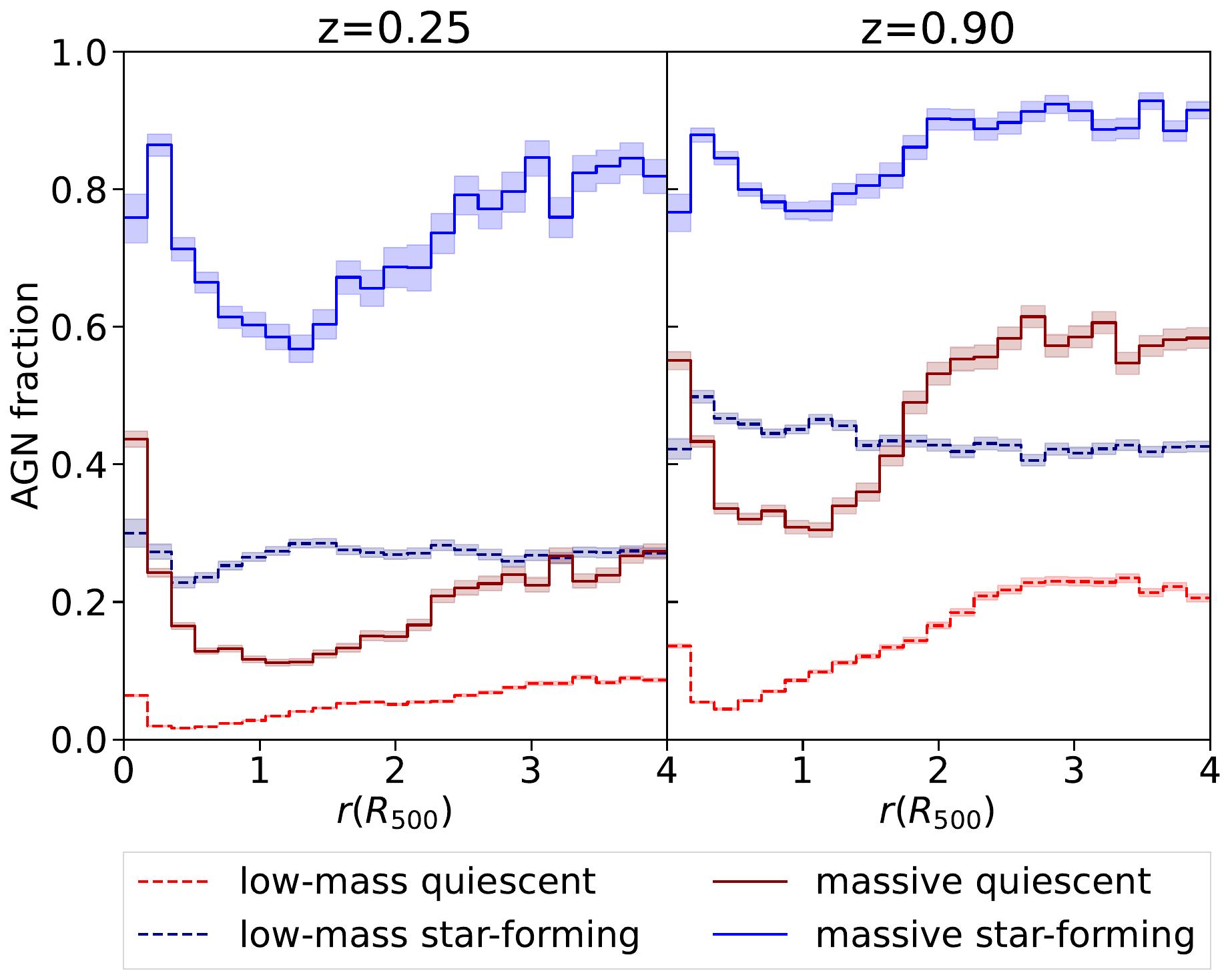}
  \caption{Fraction of galaxies containing an X-ray bright AGN as a function of $r$ at low redshift (left panel) and high redshift (right panel). The AGN fraction is plotted separately for quiescent galaxies (in red), star-forming (in blue), low-mass (dashed line) and massive galaxies (solid line). The coloured bands represent the 68 \% confidence intervals obtained with binomial statistics (see Sect. \ref{methodology}). Central BCGs are excluded.}
     \label{AGNprofile_inner}
\end{figure}

In Fig. \ref{AGNprofile_inner} we show the AGN fraction as a function of the clustercentric radius $r$ up to $4 R_{500}$ for low-mass, massive, star-forming and quiescent galaxies. Considering the most abundant low-mass quiescent galaxies (red dashed line) we notice a strong decline towards the cluster centre, which is broadly consistent with the decline of the star-forming ratio of low-mass galaxies (Fig. \ref{Asfrquiesratio_overlapping_onecolumn}). This may imply that the processes in the cluster environment affect star formation and AGN activity in low-mass galaxies similarly. Low-mass galaxies are known to be less shielded from the ICM compared to massive galaxies in our simulations (see \citealt{2019MNRAS.488.5370L}), and if the entire cold gas reservoir is removed, this is expected to affect the star formation of the whole galaxy as well as the black hole mass accretion rate. Note that the plot of the AGN fraction in quiescent galaxies allows us to infer the effects of the cluster environment on the gas content even when star formation rate is already too low to be meaningfully investigated in our simulations.

At around $2 R_{500}$ the central decline is temporarily halted, and the AGN fraction levels out. We verified that this feature becomes more pronounced if substructure members are excluded (not shown explicitly), which resembles the behaviour of the star-forming ratio in low-mass galaxies at the same $r$ (see the first panel in Fig. \ref{sfr_substructures_panels}).

At first glance, low-mass star-forming galaxies (dashed blue line in Fig. \ref{AGNprofile_inner}) show different behaviour to low-mass quiescent galaxies. At low redshift (left panel), a broad peak of the AGN fraction can be detected between $1$ and $2 R_{500}$, roughly centred around 1.3 $R_{500}$. However, before delving into potential environmental causes, we ought to carefully consider the sensitivity of AGN fraction to stellar mass, which already plays a role at the upper end of the low-mass range (see Fig. \ref{mst_ssfr}). We investigated how the observed trends change if a narrow stellar mass range ($10.6 < \log M_* (M_{odot})<10.7$) is selected instead of the entire low-mass range (not shown explicitly). We noticed that the peak at around $2 R_{500}$ becomes less pronounced, and the central decline resembles the behaviour of quiescent galaxies. The occurrence of mass segregation of star-forming galaxies is not surprising; in Sect. \ref{star_formation} we demonstrated that the low-mass galaxies are quenched more easily than massive galaxies, thus the star-forming population in the inner regions consists of more massive galaxies, hosting more massive and more active black holes. The central decline below $1 R_{500}$, however, is still a consequence of environmental effects. It tells us that the AGN activity in low-mass galaxies is suppressed in the cluster centre, even if the galaxies retain enough gas to form stars. 

Massive galaxies (solid lines in Fig. \ref{AGNprofile_inner}) display somewhat different behaviour than low-mass galaxies. We can see a gradual decline of the AGN fraction towards the inner regions, with a minimum between 1 and 2 $R_{500}$, followed by a drastic increase in the centre. This shape is found in massive star-forming and quiescent galaxies and is not caused by different stellar masses (we verified that the trend persists even if a narrow $M_*$ range is selected). When comparing to Fig. \ref{Asfrquiesratio_overlapping_onecolumn} we can see that the AGN fraction does not match the star-forming ratio; The minimum of AGN activity around $1 R_{500}$ is contrasted by significantly elevated star formation in massive galaxies.

While this may hint at a possible causal connection between the cessation of AGN activity and elevated star formation, it is more likely an indicator that the AGN activity and star formation in massive galaxies are not affected by the same processes to the same extent. For instance, we found that, unlike star-forming ratio (Fig. \ref{sfr_substructures_panels}), the behaviour of AGN fraction is qualitatively the same in members of substructures and single galaxies that orbit the cluster alone - all galaxies show diminished AGN activity around $1 R_{500}$. The differences between star formation and AGN activity are not inconceivable since, apart from the availability of cold gas (and its compression) that drives star formation, AGN activity also requires the loss of angular momentum and transport of the gas towards the galactic centre. One possible explanation of the trends would be that the abundance of galaxy mergers and interactions on the cluster outskirts keeps the AGN fraction (and star-forming ratio) high. Inside the virial radius, merging activity seizes due to higher velocity dispersion, and AGN activity drops. On the other hand, star formation in massive galaxies is triggered after infall, causing the peak of the star-forming fraction, and drops again when the galaxy approaches the centre ($r< 1 R_{500}$) (discussed in section \ref{star_formation} and investigated in \citealt{2019MNRAS.488.5370L}).

We also tried to identify the possible reason for the increase of AGN fraction in massive galaxies below $1 R_{500}$, seen in Fig. \ref{AGNprofile_inner}. When examining galaxies in narrow stellar mass ranges, we noticed that their median black hole mass increases towards the cluster centre at low $r$ (most drastically below $1 R_{500}$). Higher black hole mass then leads to higher AGN fraction in massive galaxies, as discussed in section \ref{AGNactivity}. We verified that the black hole mass increase does not result from spurious \textsc{Subfind} $M_*$ determination. It remains even if maximum circular velocity is used as a proxy for stellar mass. We propose two reasons for the increasing black hole mass towards the cluster centre. One is faster black hole growth relative to stellar mass growth. When examining the ratio between black hole mass accretion rate and star formation rate, we found an increase below 0.5 $R_{500}$ (i.e., star formation is, on average, more suppressed than black hole mass accretion rate, which increases in the centre). Over time, this increase would result in even higher black hole masses; however, mass accretion rates found in our simulation cannot fully account for the high black hole mass increase on reasonable time scales. The second possible reason is tidal stripping. Galaxies in the inner regions can lose both dark matter and stellar mass due to tidal interactions with cluster members and cluster potential \citep[e.g.,][]{2023MNRAS.tmp.3185M,2016ApJ...833..109S,2012MNRAS.427.1024I,2010MNRAS.406..729S,2006MNRAS.366..429R}. A decrease in stellar mass can then be reflected in higher black hole mass at constant $M_*$. We found an indication of the presence of tidal stripping by examining the dark matter halo to stellar mass ratio. Since dark matter halos are more extended than stellar components, they get stripped more effectively \citep[e.g.,][]{2016ApJ...833..109S,2013MNRAS.429.1066S,2008ApJ...673..226P}. We found that dark matter halo mass (relative to stellar mass) decreases drastically towards the cluster centre. Dark matter halo mass of galaxies with constant $M_*$ decreased by a factor of $\sim3$ when going from 3 to $0.5 R_{500}$. To evaluate the contribution of both effects to the increase of median black hole mass at low $r$, a comprehensive study of the time evolution of a large number of galaxies would be required, which is beyond the scope of this work. We should also keep in mind that some of those central black holes may, in reality, be accreting with low radiative efficiency and might not be considered X-ray bright.

\begin{figure}
\centering
\includegraphics[width=\textwidth*19/40]{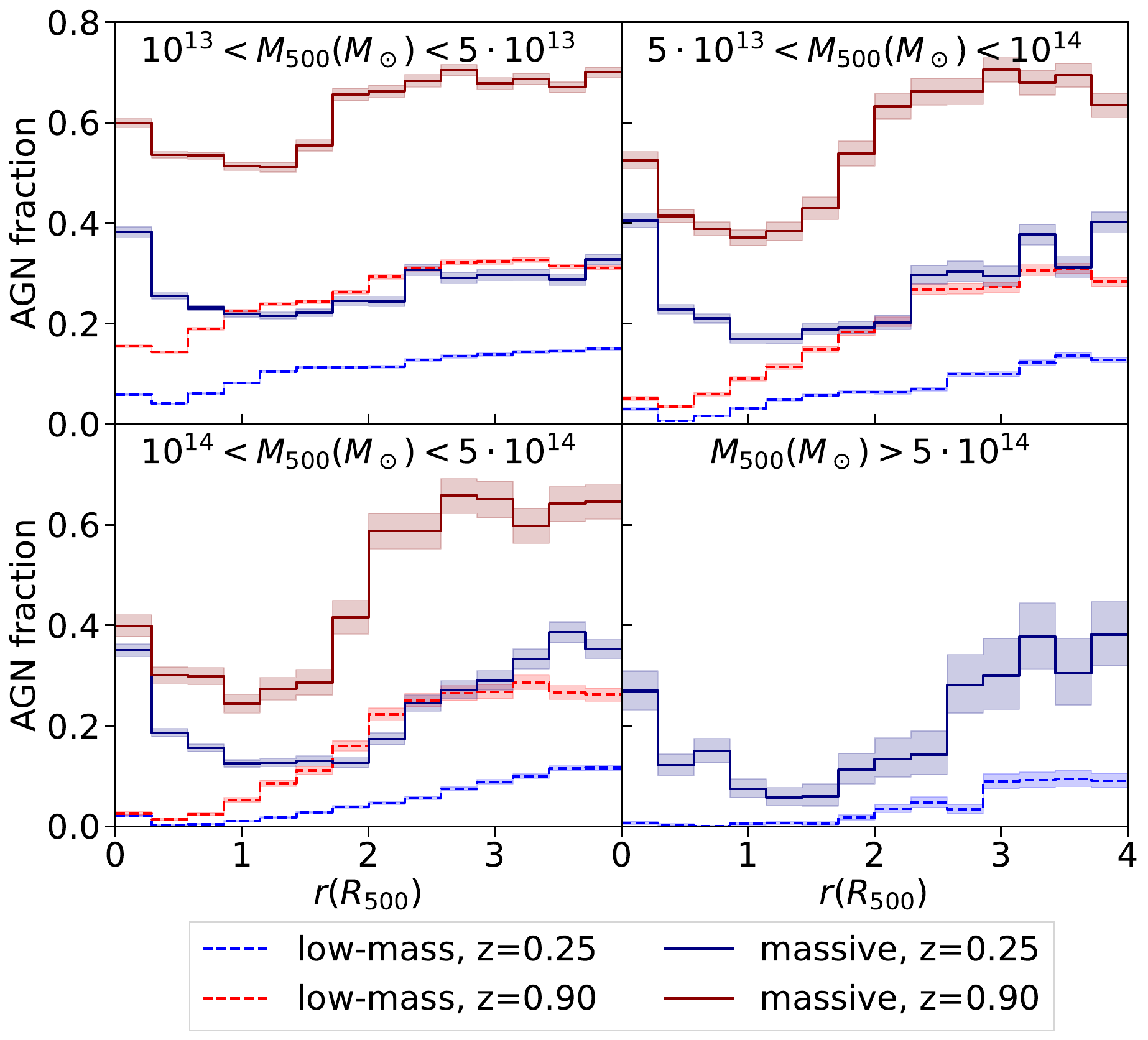}
  \caption{AGN fraction as a function of $r$, shown for low-mass (dashed line) and massive galaxies (solid line) at low redshift (in blue) and high redshift (in red). Each panel represents the AGN fraction in clusters of different mass $M_{500}$ range. High-redshift massive clusters are omitted due to their low abundance and thus high statistical uncertainty. The coloured bands represent the 68 \% confidence intervals obtained with binomial statistics (see Sect. \ref{methodology}). Central BCGs are excluded. }
     \label{AGNprofili_M500}
\end{figure}

So far, we have considered all clusters in the simulation with $M_{500}>10^{13}M_{\odot}$ - the sample dominated by low-mass clusters. We also investigated whether the trends shown so far are preserved in more massive clusters. In Fig. \ref{AGNprofili_M500} we show the AGN fraction as a function of $r$ in low-mass and massive galaxies in clusters in four $M_{500}$ ranges. Qualitatively, the behaviour does not depend on $M_{500}$, except that the suppression of AGN activity at low $r$ is stronger in massive clusters (i.e., the AGN fraction at low $r$ is lower). There is also a mild indication that the delay in the AGN fraction drop, found in the profile of low-mass galaxies around $2 R_{500}$, is more pronounced in low-mass clusters below $10^{14} M_\odot$ (upper panels). In clusters between $10^{14}$ and $5 \times 10^{14} M_\odot$ (bottom left panel), the decline in the central regions is smoother.  

Another question that may arise is whether the trends change if the AGN luminosity threshold is increased and only the brightest AGNs are selected. When the luminosity threshold was raised from $10^{42}$ erg/s to $10^{43}$ erg/s we found no significant qualitative difference compared to radial profiles shown in Fig. \ref{AGNprofile_inner}. Raising the threshold even more proved to be meaningless due to the small AGN sample size.

\subsection{AGN fraction on the cluster outskirts}
\label{AGN_fraction_at_large_r}

\begin{figure}
\centering
\includegraphics[width=\textwidth*19/40]{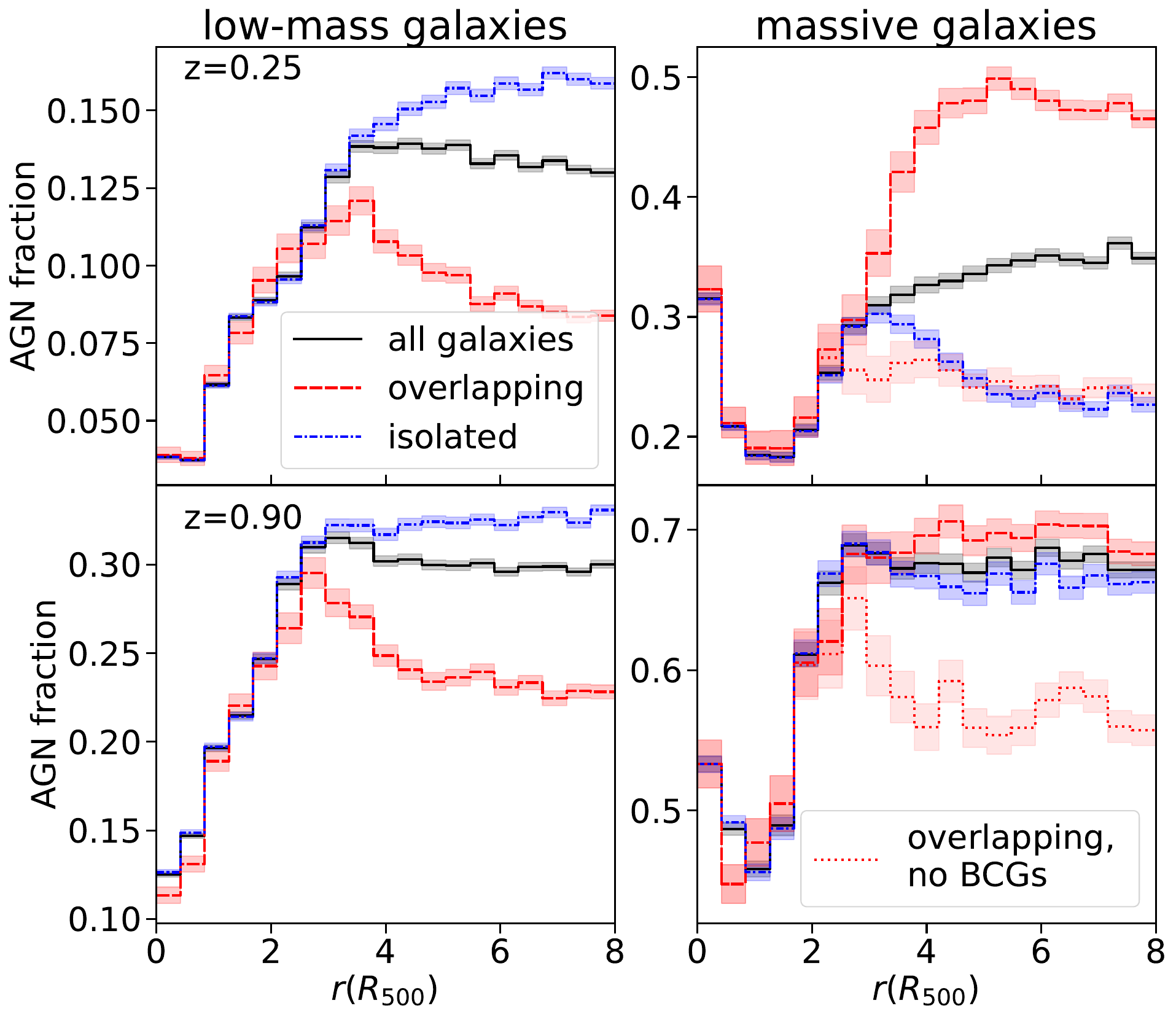}
  \caption{Fraction of galaxies hosting an X-ray bright AGN as a function of clustercentric distance $r$. The fraction of low-mass galaxies ($10.15 < \log M_*(M_\odot)<11$) is shown on the left and of the massive galaxies ($\log M_*(M_\odot)>11$) on the right. Results are shown for low redshift (upper panels) and high redshift (lower panels) and for the overlapping and isolated regions ($r_{ov}=6 R_{500}$, see Sect. \ref{methodology}). On the right panels, the dotted line represents the massive overlapping populations without the BCGs of the neighbouring clusters. Central BCGs are excluded in all plots. The coloured bands represent the 68 \% confidence intervals obtained with binomial statistics (see Sect. \ref{methodology}). }
\label{AGNprofile_overlapping_column}
\end{figure}

In Fig. \ref{AGNprofile_overlapping_column} we plot the AGN fraction as a function of $r$ in extended cluster regions up to $8 R_{500}$. As outlined in Sect. \ref{influence_of_neighbours} and already done with the star-forming ratio in Fig. \ref{Asfrquiesratio_overlapping_onecolumn}, we take into account the large-scale surroundings of the clusters and separately consider overlapping and isolated regions. The profile of low-mass galaxies (left panels) is almost identical to the profile of the star-forming ratio in Fig. \ref{Asfrquiesratio_overlapping_onecolumn} with central suppression of AGN activity, which is reflected in a drop of AGN activity at large $r$ in the overlapping population. The peak of AGN fraction between $2$ and $4 R_{500}$ can be explained with geometric overlaps, as was demonstrated in Fig. \ref{overlaps_MC} and does not require an excess of AGN activity on the outskirts of individual low-mass clusters. This peak in the overlapping regions is also the reason for the subtle peak in the combined profile at high redshift (solid black line on the lower left panel) at around $3 R_{500}$; it is absent in the profile of isolated regions (blue dash-dotted line). 

Considering massive galaxies (right panels), we can notice similar behaviour in the overlapping population, especially when the contribution of the neighbouring BCGs is removed (red dotted line). But interestingly, the isolated population also clearly shows an excess of AGN activity on the outskirts, at $\sim 3 R_{500}$. This excess remains even if we tighten the definition of the isolated regions and only keep single massive galaxies that do not belong to any group with multiple galaxies (and not just clusters above our selection threshold). Hence, this excess most likely shows that the outskirts of clusters are a particularly suitable environment for AGN activity, as was reported by observational studies by \cite{2019A&A...623L..10K}.
We should also remember that at the same $r$, signs of an excess of the star-forming ratio were found (Fig. \ref{Asfrquiesratio_overlapping_onecolumn}), meaning that unlike in the inner regions, the conditions in the outer regions may trigger both AGN activity and star formation. We expect this excess of AGN activity to be particularly noticeable in relaxed clusters in isolation that do not contain massive groups in their vicinity. If the profiles of all clusters are combined (black solid line), the excess of AGN activity is not discernible anymore at low redshift (upper right panel); the profiles of overlapping and isolated regions have their peaks at different $r$, and the physical excess in the isolated population is counterbalanced by the rapid ascent of the AGN fraction in the overlapping population (probably due to geometric overlaps). Tentative signs of the AGN excess can still be identified in the combined profile at high redshift (bottom right panel) since the peak due to geometric overlaps coincides with the excess in the isolated population. Furthermore, we verified that the profiles are qualitatively similar in massive clusters ($M_{500}>10^{14}M_\odot$) with a few minor differences, e.g., the peak due to geometric overlaps in massive galaxies in the overlapping regions is reached at lower $r$ (around 3 $R_{500}$) compared to all clusters (see the red line in the upper right panel of Fig. \ref{AGNprofile_overlapping_column}).

\subsection{AGN fraction and local density}
\label{agn_and_local_density}

\begin{figure*}
\centering
\includegraphics[width=\textwidth*9/11]{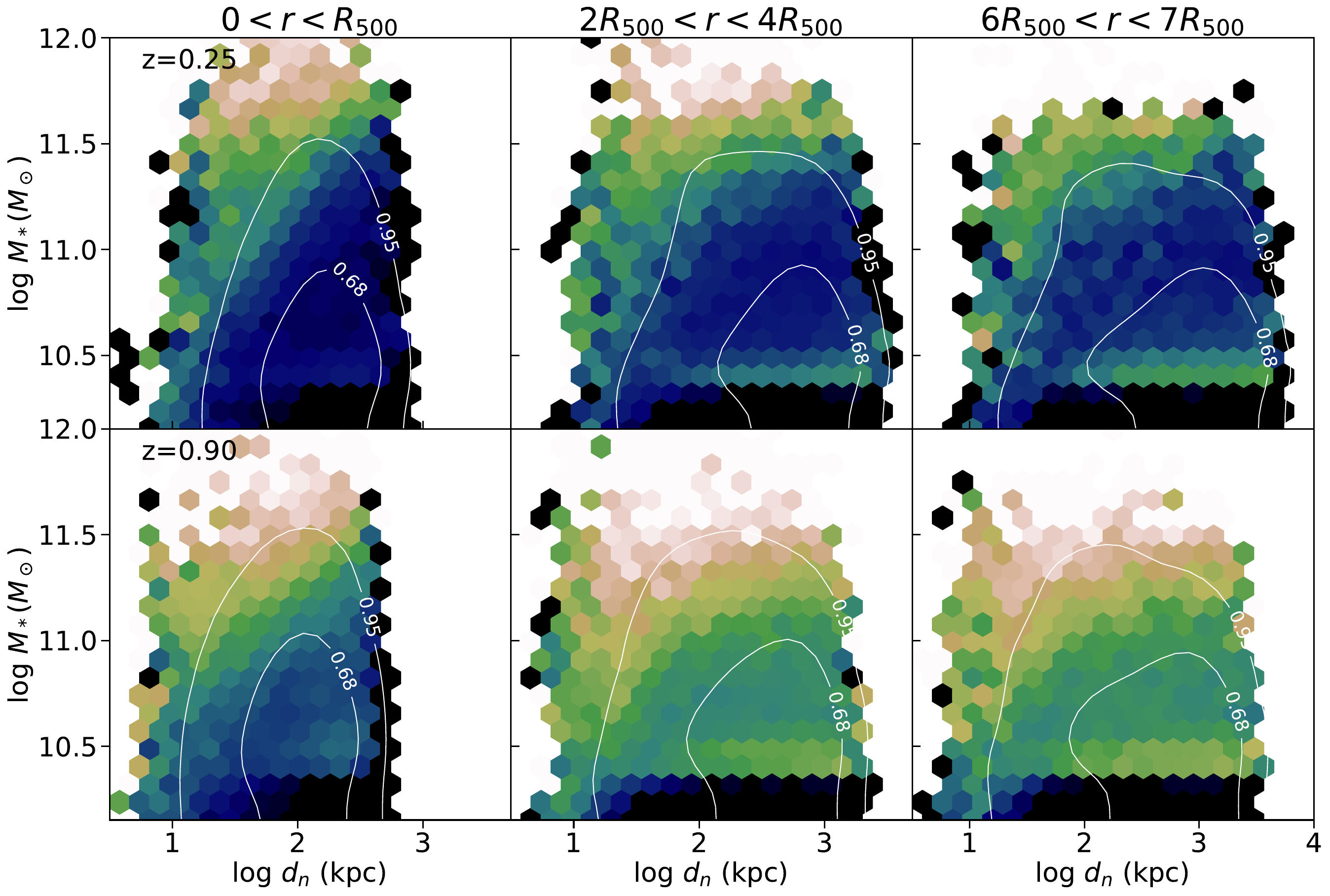}
\includegraphics[width=\textwidth*9/11]{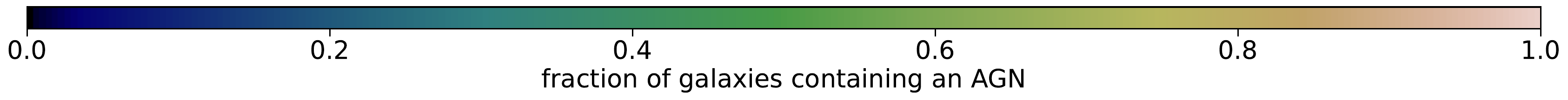}
\caption{Fraction of galaxies hosting an X-ray bright AGN as a function of the distance to the nearest neighbour $d_n$ and stellar mass $M_*$. The contours represent the distribution of galaxies. Results are displayed for different radial ranges (different columns). In the rightmost column, only isolated regions were selected to remove the contributions of the neighbouring clusters. The upper row shows the fraction at low redshift and the lower row at high redshift. The bins where the $68\%$ binomial confidence interval of the AGN fraction is wider than 0.2 are not displayed.}
\label{frac_dnb_mst_isolated_nobcg}
\end{figure*}

As in Sect. \ref{sfr_local_density} we tried to disentangle the effects of the local density on AGN activity from radial dependence. In Fig. \ref{frac_dnb_mst_isolated_nobcg}, we plot the AGN fraction as a function of the distance to the nearest neighbour $d_n$ and stellar mass $M_*$ at low and high redshift in the inner cluster regions (left panels), on the outskirts (middle panels) and in the field (right panels). We can see that the AGN fraction is drastically higher in a locally denser environment (low $d_n$). We have already partially discussed this trend in section \ref{radial_dependence_of_AGN_fraction}. In the innermost regions, the expected $d_n$ of galaxies is highly correlated with $r$. Thus, the increase of black hole mass and the AGN fraction with decreasing $r$, shown in Fig. \ref{radial_dependence_of_AGN_fraction}, is also reflected in the rise of AGN fraction with local density in the left panels of Fig.\ref{frac_dnb_mst_isolated_nobcg}. However, we verified that this trend holds more generally. We verified that the median black hole mass rises with local density not only in the cluster centre but also at larger $r$. This is reflected in the increased AGN activity at small $d_n$ (see middle and right panels in figure \ref{frac_dnb_mst_isolated_nobcg}).  We performed similar tests as in section \ref{radial_dependence_of_AGN_fraction} and found signs of an increase of black hole mass accretion relative to star formation rate, as well as a decrease in the ratio between dark matter halo and stellar mass in very dense environments at larger $r$. Thus, we can argue that the same mechanisms, as discussed in section \ref{radial_dependence_of_AGN_fraction} (faster black hole growth relative to star formation, tidal stripping), are also applicable in dense regions at high $r$. Aside from the black hole mass increase, we cannot rule out other environmental factors. Interestingly, the general behaviour and the increasing trend with local density, especially in massive galaxies with $d_n > 50 \mathrm{kpc}$, resembles the behaviour of the star-forming ratio in Fig. \ref{dnb-sfrquiesratio-narrowregions_single}, which does not directly depend on the black hole mass. Furthermore, we verified that, similarly to the star-forming ratio, the drastic $d_n$ dependence of the AGN fraction is more pronounced in the members of massive substructures with multiple galaxies. We found that most galaxies with $d_n$ below a few hundred kpc are substructure members.

Despite the strong influence of $d_n$ on star-forming ratio and AGN fraction, we verified that it cannot fully explain all trends discussed in sections \ref{star_formation} and \ref{radial_dependence_of_AGN_fraction}. We investigated radial profiles of AGN fraction and the star-forming ratio of massive galaxies in a narrow range of local densities ($1.8<\log d_n(\mathrm{kpc})<1.9$, not shown explicitly). We found that both the decline of AGN activity inside the cluster, observed in Fig. \ref{AGNprofile_inner}, and the increase in the star-forming ratio, observed in Fig. \ref{Asfrquiesratio_overlapping_onecolumn}, can still be identified with similar amplitudes if $d_n$ is constant. However, the central rise of the AGN fraction is almost absent, pointing to the importance of the increasing black hole mass with local density, discussed in the previous paragraph.

\subsection{Comparison with observational studies}
\label{comparisson_with_xray}
\begin{figure*}
\centering
\includegraphics[width=\textwidth*9/10]{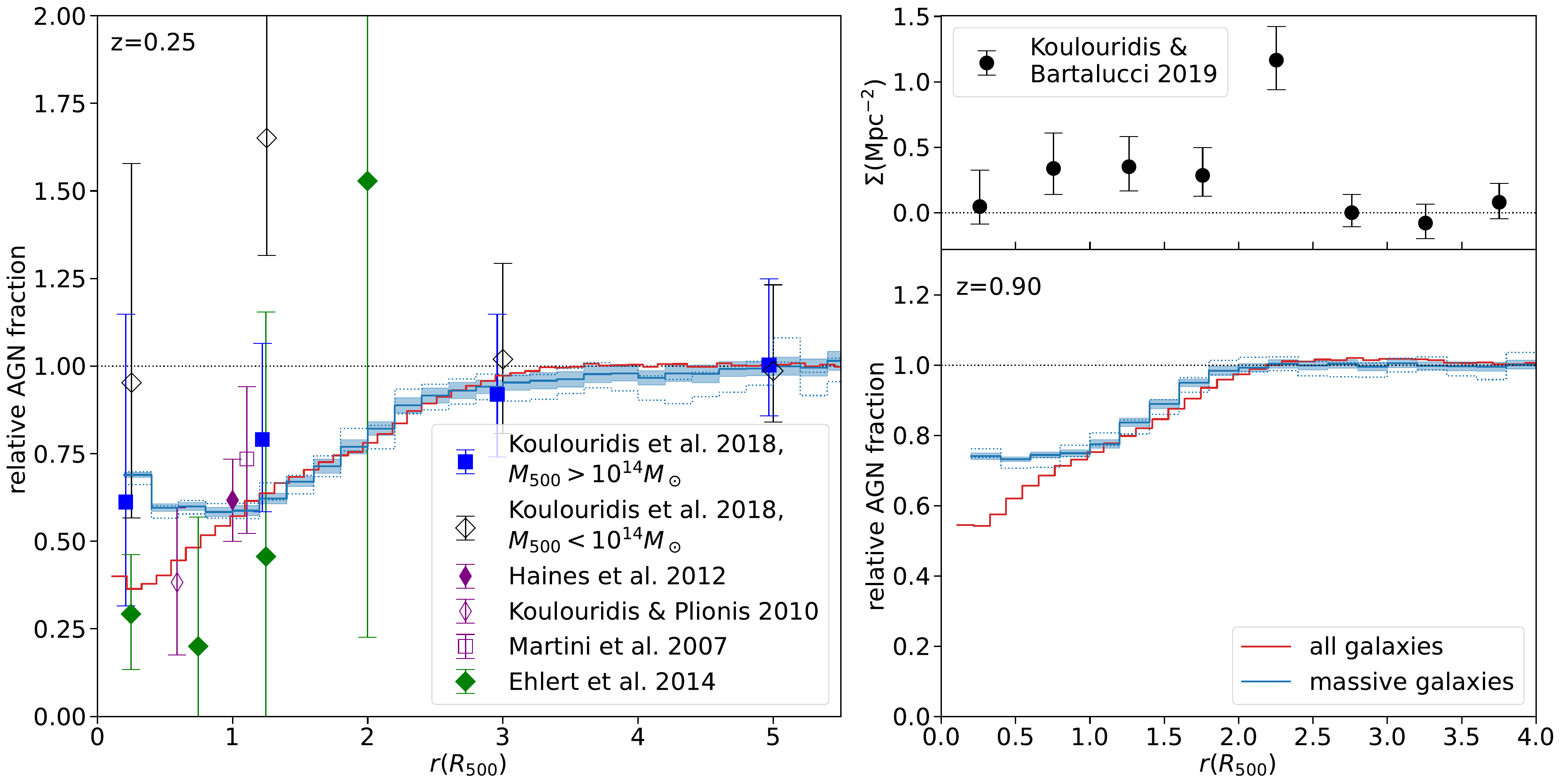}

\caption{Fraction of galaxies containing an X-ray bright AGN, divided with the background fraction, as a function of \textit{projected} clustercentric distance $r$. {\it Magneticum} results are shown at low redshift (left panel) and high redshift (bottom right panel) for all galaxies with $\log M_*(M_\odot)>10.15$ (in red) and only massive galaxies ($\log M_*(M_\odot)>11$, in blue). The projection depth is 7.6 Mpc. The plot represents the mean profile of many lines of sight and the coloured interval $\pm$ standard deviation of different lines of sight. Dotted lines mark the $68\%$ binomial confidence interval when only one line of sight is considered. The confidence intervals of low-mass galaxies are too narrow to be meaningfully displayed. The centralmost bins, influenced by the BCGs, are omitted. The background AGN fraction is defined as the mean fraction in the annulus between $4.5$ and $6$ $R_{500}$. The fraction at low redshift (left panel) is compared with results of \cite{2014MNRAS.437.1942E} and \cite{2018A&A...620A..20K}, from which we also obtained the results of \cite{2007ApJ...664..761M}, \cite{2010ApJ...714L.181K} and \cite{2012ApJ...754...97H}. In the upper right panel, we plot the quantity $\Sigma$ from figure 1 in \cite{2019A&A...623L..10K}.  The data points from the literature are
placed at the centre of radial bins and are shown without the radial errorbars.}
\label{radial_profile_literature_2}
\end{figure*}

We compared the radial profiles of the AGN fraction in our simulations with the results of several X-ray studies, which are summarized in table \ref{tab:literature_data}. In Fig. \ref{radial_profile_literature_2} we plot their results together with the results of the {\it Magneticum} simulations. We separately plot the AGN fraction in low-mass galaxies and massive galaxies at low and high redshift as a function of projected distance $r$ with projection depth 7.6 Mpc\footnote{The projection distance 7.6 Mpc matches 6 virial radii of the most massive
cluster in the simulation at high redshift.} as opposed to the three-dimensional distance that was used in Fig. \ref{AGNprofile_inner} and \ref{AGNprofile_overlapping_column}. The studies used different redshift ranges, cluster mass ranges, and the galaxy and AGN selection criteria (see Table \ref{tab:literature_data}), which makes direct comparison with our results challenging - our study indicated that the definition of the galaxy sample and the different selections in terms of mass and geometrical configuration might impact the resulting trends. By doing an order of magnitude estimate, it can be seen that many studies in table \ref{tab:literature_data} use stellar mass limit close to $10^{10} M_\odot$, which corresponds to the low-mass galaxies in this work. However, this threshold is critically close to the black hole seeding value - more massive galaxies are better resolved in our simulations and might be more suitable for comparison with observations. 

Almost all studies in Fig. \ref{radial_profile_literature_2} report a significant decrease in AGN activity in the inner cluster regions \citep{2018A&A...620A..20K, 2014MNRAS.437.1942E, 2012ApJ...754...97H, 2010ApJ...714L.181K, 2007ApJ...664..761M}, which is consistent with our findings. As for other features discussed in this section, they are difficult to discern in Fig. \ref{radial_profile_literature_2} since they only appear in certain populations of galaxies and since projection further blurs any radial dependence, but we can nonetheless try to identify some similarities with observations. The profile of low-mass clusters from \cite{2018A&A...620A..20K} (where the galaxy magnitude selection criteria, roughly correspond to our low-mass galaxies) shows an absence of strong central decline and an excess of AGN activity between $1$ and $2 R_{500}$. This is in qualitative agreement with the profiles of low-mass galaxies in low-mass clusters found in the simulations (top left panel of Fig. \ref{AGNprofili_M500}), where the central decline is shallower compared to massive clusters (bottom panels). Although no excess is found between $1$ and $2 R_{500}$ in Fig. \ref{AGNprofili_M500}, there is a plateau of AGN activity which could be interpreted as an excess with respect to the overall decline. This plateau is less apparent in massive clusters.

Particularly interesting are the results of \cite{2019A&A...623L..10K}. In the top right panel, we plot the quantity $\Sigma$ from figure 1 in \cite{2019A&A...623L..10K}, which represents the total surface density of X-ray point sources in excess of the field value, divided by the optical galaxy profile in high-redshift clusters (around $z=1$). The plot clearly shows an excess of AGN activity between $2 R_{500}$ and $2.5 R_{500}$. Note that at approximately this distance, we find a slight excess of AGN activity in massive galaxies at high redshift (bottom right panel of figure \ref{AGNprofile_overlapping_column}), where the peaks in isolated regions (due to the environmental conditions) and in the overlapping regions (due to geometric overlaps) align. Signs of excess are also present in low-mass galaxies (bottom left panel of figure \ref{AGNprofile_overlapping_column}), which are closer to the mass range used by \cite{2019A&A...623L..10K}. It should be, however, stressed that $\Sigma$ is not equivalent to the AGN fraction we used. 
Different approaches adopted by both works prevent a direct quantitative comparison of our results. Whereas the existence of the peak and its cluster-centric distance $r$ is found in both works the amplitudes of the peaks cannot be meaningfully compared.

\begin{table*}
\caption{Redshift range, cluster mass range, AGN luminosity $L_X$ (or flux $F_X$) of the AGNs, and magnitudes of optically selected galaxies from several studies, shown in Fig. \ref{radial_profile_literature_2}. Refer to original papers and references therein for descriptions of the magnitude thresholds $R$, $M^*_i, M_k$ and $m_r$, and details about the selected clusters where their mass was not given explicitly.}           
\label{tab:literature_data}    
\centering          

\begin{tabular}{ c c c c c }
Study & $z$ range & $M_{500} (M_\odot)$ & AGN X-ray selection & galaxy selection \\

\hline
\hline

Koulouridis 2019& 0.93-1.13 & \begin{tabular}{@{}c@{}} $5 \cdot 10^{14} -$\\$9 \cdot 10^{14}  $\end{tabular} &  \begin{tabular}{@{}c@{}}$L_{X}(0.5-8 \mathrm{keV})>$\\$3 \times 10^{42}$ erg s $^{-1}$\end{tabular} &  \begin{tabular}{@{}c@{}}$R<23,$\\ SuprimeCam R-band, \\ $M_* \gtrsim 10^{10} M_\odot$\end{tabular}\\
\hline
Koulouridis 2018& 0.1-0.5 & \begin{tabular}{@{}c@{}} $ 10^{13} -$\\$5 \cdot 10^{14}  $\end{tabular} &  \begin{tabular}{@{}c@{}}$L_{X}(0.5-10 \mathrm{keV})>$\\$10^{42}$ erg s $^{-1}$\end{tabular} &  \begin{tabular}{@{}c@{}}$-23.75 \lesssim M^*_i \lesssim -20.75$\end{tabular}\\
\hline

Ehlert 2014 & 0.2-0.7 & \begin{tabular}{@{}c@{}}$4.7 \cdot 10^{14} -$\\$2.2 \cdot 10^{15} $\end{tabular} &  \begin{tabular}{@{}c@{}}$F_{X}(0.5-8.0 \mathrm{keV})>$\\ $10^{-14} \operatorname{erg~cm}^{-2} \mathrm{~s}^{-1}$\end{tabular} &  \begin{tabular}{@{}c@{}}$R<23,$\\ SuprimeCam R-band, \\ $M_* \gtrsim 10^{10} M_\odot$\end{tabular}\\
\hline

Haines 2012 & 0.15-0.3 & $ \gtrsim5 \cdot 10^{14} $ &  \begin{tabular}{@{}c@{}}$L_{X}(0.3-7 \mathrm{keV}) \gtrsim$\\$ 10^{42}$ erg s $^{-1}$\end{tabular}&  \begin{tabular}{@{}c@{}}$M_k<-23.1$\end{tabular}\\
\hline

Koulouridis 2010 & 0.07-0.28 &  &  \begin{tabular}{@{}c@{}}$L_{X}(0.5-8 \mathrm{keV}) >$\\$ 10^{42}$ erg s $^{-1}$\end{tabular}&  \begin{tabular}{@{}c@{}}$m_{r}^{*}-0.5<m_{r}<m_{r}^{*}+0.5$\end{tabular}\\
\hline

Martini 2007 & 0.06-0.31 &  &  \begin{tabular}{@{}c@{}}$L_{X} > 10^{42}$erg s $^{-1}$ and\\$L_{X} > 10^{41}$erg s $^{-1}$\end{tabular}&  \begin{tabular}{@{}c@{}}$M_R < -20$\end{tabular}\\

\end{tabular}
\end{table*}

\section{Summary \& Conclusions}
\label{conclusion}
In this work we investigated the properties of cluster galaxies in the {\it Magneticum} suite of cosmological hydrodynamical simulations, with particular emphasis on the AGN activity and levels of star formation in cluster galaxies as a function of their environment. For our analysis we used a cosmological box spanning over 640 comoving Mpc/$h$ which provided a large sample of clusters and galaxies for a statistically meaningful analysis. We mainly considered clusters with $M_{500}>10^{13}M_\odot$ and galaxies with $\log M_*(M_\odot)> 10.15$ split in two mass ranges and investigated whether they contained AGNs with X-ray luminosity above $10^{42}$ erg/s in 0.5-10 keV band. We identified several trends in certain populations of galaxies and compared their behaviour with X-ray studies. We also provided some possible interpretations of the observed trends, with our main conclusions summarized as follows. 

   \begin{enumerate}
      \item  By splitting the galaxies into overlapping and isolated populations based on their distance to the neighbouring clusters, we explored the influence of the large-scale surroundings on the radial dependence of galaxy properties in clusters. We predicted the influence of the neighbouring clusters by generating a set of random points around each cluster and in each point, estimating the contribution of all (central and neighbouring) clusters based on their distance from that point. We computed the average distance of galaxies to their parent clusters at each point, which we use as a proxy for their properties. We found that in the overlapping population, the galaxies are, on average, the furthest away from any cluster if they are between $r=3R_{500}$ and $4 R_{500}$ away from the central cluster (if the overlapping distance of $6R_{500}$ is chosen). If the distance $r$ increases even more, the inner regions of neighbouring clusters become increasingly important. This prediction was tested against the radial profile of the fraction of galaxies hosting an AGN (AGN fraction) and of the ratio between the number of star-forming and quiescent galaxies (star-forming ratio) in the overlapping population, and it was indeed found that they are the highest at around the same $r$; the shape of the profile of the overlapping populations is likely a consequence of geometric overlaps between cluster regions and does not require additional physical effects which would promote star formation or AGN activity on the cluster outskirts. Furthermore, the overlaps may conceal some features that are clearly discernible in the isolated population and should thus be carefully accounted for when the properties of clusters and cluster galaxies are studied up to very high radii (above $\sim3 R_{500}$).

      \item We demonstrate how the AGN activity in galaxies with different stellar masses reflects the implementation of supermassive black holes in the simulations; recently seeded black holes produce a peak of AGN activity in galaxies with $\log M_* (M_\odot) \approx 10.4$, rapidly quenching their host and reducing the AGN fraction. The strength of the peak of AGN activity in low-mass galaxies ($10.15 < \log{M_*( M_\odot)} < 11$) depends strongly on the global environment (clustercentric radius). In massive galaxies on the other hand ($\log M_* (M_\odot) > 11$), the AGN fraction is decisively influenced by the black hole mass (and consequently stellar mass).
    
   \item The AGN fraction and star-forming ratio in low-mass galaxies rapidly decline in the inner cluster regions, suggesting a major role of processes such as ram pressure stripping, strangulation and other environmental quenching mechanisms in the depletion of the cold gas reservoir in cluster galaxies. A similar central decline is also reported in observational studies of star formation \citep[e.g.,][]{2014ApJ...783..136C, 2012MNRAS.424..232W} and X-ray bright AGNs \citep{2018A&A...620A..20K, 2014MNRAS.437.1942E, 2012ApJ...754...97H, 2010ApJ...714L.181K, 2007ApJ...664..761M} in galaxy clusters. 
   
   \item Massive galaxies, on the other hand, show an increase in the star-forming ratio towards the inner cluster regions, which peaks at around $R_{500}$, followed by a steep decline in the innermost regions. This is consistent with a brief increase of star formation after the in-fall in massive galaxies, reported by \cite{2019MNRAS.488.5370L} using the same simulations. The peak is especially prominent in substructure members, that are more likely in-falling. This trend is in stark contrast to the AGN fraction in massive galaxies, which exhibits a decline of AGN activity in the inner regions with a sharp rise in the cluster centre below $R_{500}$, regardless of the substructure membership. The difference indicates that the processes in the cluster environment do not have the same effect on star formation and AGN activity in massive galaxies.

   \item  We found that an increase of AGN activity in massive galaxies below $1R_{500}$ is a consequence of increasing black hole masses relative to stellar mass in denser local environments. The increase is likely a consequence of stellar mass stripping or higher black hole mass accretion relative to star formation rate.

   \item We explored the influence of the local density (traced with the distance to the nearest neighbour $d_n$) on the star-forming ratio and the AGN fraction and found that low-mass galaxies in the field have the highest star-forming ratio if they are in environments with very low local density (large $d_n$), indicating the role of environmental quenching. Massive galaxies in the field exhibit opposite behaviour - their star-forming ratio is the lowest in low-density environments, potentially due to mass quenching. Inside the clusters, higher local density generally means a higher star-forming ratio and AGN fraction regardless of stellar mass. The local density dependence is especially strong in substructure members.

    \item While local density could, in principle, play a role in $r$ dependence as it is highly correlated with $r$, we verified that it cannot account for most of the observed trends. For example, central AGN suppression and the increase of the star-forming ratio in massive galaxies remain even if a narrow $d_n$ range is selected, indicating the impact of the global cluster environment. They persist even if a narrow stellar mass range is chosen, thus ruling out the decisive influence of mass segregation. The increase of AGN fraction below $1 R_{500}$, on the other hand, becomes almost absent if the local density is kept constant. This further proves the role of increasing black hole mass with local density.

   \item We report increased AGN activity on the cluster outskirts. The peak of AGN fraction at around $3 R_{500}$ is discernible in the isolated population of galaxies. However, when all galaxies are selected, this trend is contaminated by overlaps with neighbouring cluster regions, albeit it still remains marginally visible at high redshift. The increase in AGN activity on the cluster outskirts may correspond to the trends found in X-ray study by \cite{2019A&A...623L..10K}.

    \item We found that the expected value of specific star formation rate sSFR in low-mass star-forming galaxies remains constant despite the declining star-forming ratio in the inner regions, whereas sSFR of massive star-forming galaxies roughly traces the star-forming ratio and is lower in the cluster centre.

   \end{enumerate}

In this work, we demonstrated that despite the complexity of the underlying physical processes, which poses a numerical and computational challenge for simulations, the {\it Magneticum} simulations capture 
several observed trends and can thus shed some light on the underlying physical mechanisms. The trends reported in this work should be investigated in more detail in future studies to better constrain their physical origin by a complementary kinematic analysis, for instance. This would allow identifying the infalling population and recent gravitational interactions, providing further insight into the detailed physical processes at work. 

\begin{acknowledgements}
GR acknowledges the support from the Ad futura Programme of the Public Scholarship, Development, Disability and Maintenance Fund of the Republic of Slovenia and from the Slovenian national research agency ARRS through grant N1-0238. KD acknowledges support by `BiD4BEST' - European Innovative Training Network (ITN) funded by the Marie Sk\l{}odowska-Curie Actions (860744) in Horizon 2020. VB acknowledges partial support by the Deutsche Forschungsgemeinschaft (DFG, German Research Foundation) --- 415510302, and by the Excellence Cluster ORIGINS which is funded by the DFG under Germany's Excellence Strategy – EXC-2094 – 390783311. DF acknowledges financial support
from the Slovenian Research Agency (research core funding no.
P1-0188).
The calculations for the hydrodynamical simulations were carried out at the Leibniz Supercomputer Center (LRZ) under the project pr83li ({\it Magneticum}). We would also like to thank Elias Koulouridis for kindly providing details about the \cite{2019A&A...623L..10K} paper and the anonymous referee whose comments significantly enhanced the quality of this paper.

\end{acknowledgements}

%
\bibliographystyle{aa} 
\bibliography{bib}{} 

\begin{thebibliography}{95}
\expandafter\ifx\csname natexlab\endcsname\relax\def\natexlab#1{#1}\fi

\bibitem[{{Abadi} {et~al.}(1999){Abadi}, {Moore}, \& {Bower}}]{1999MNRAS.308..947A}
{Abadi}, M.~G., {Moore}, B., \& {Bower}, R.~G. 1999, \mnras, 308, 947

\bibitem[{{Arth} {et~al.}(2014){Arth}, {Dolag}, {Beck}, {Petkova}, \& {Lesch}}]{2014arXiv1412.6533A}
{Arth}, A., {Dolag}, K., {Beck}, A.~M., {Petkova}, M., \& {Lesch}, H. 2014, arXiv e-prints, arXiv:1412.6533

\bibitem[{{Balogh} {et~al.}(1998){Balogh}, {Schade}, {Morris}, {Yee}, {Carlberg}, \& {Ellingson}}]{1998ApJ...504L..75B}
{Balogh}, M.~L., {Schade}, D., {Morris}, S.~L., {et~al.} 1998, \apjl, 504, L75

\bibitem[{{Beck} {et~al.}(2016){Beck}, {Murante}, {Arth}, {Remus}, {Teklu}, {Donnert}, {Planelles}, {Beck}, {F{\"o}rster}, {Imgrund}, {Dolag}, \& {Borgani}}]{2016MNRAS.455.2110B}
{Beck}, A.~M., {Murante}, G., {Arth}, A., {et~al.} 2016, \mnras, 455, 2110

\bibitem[{{Bekki} \& {Couch}(2003)}]{2003ApJ...596L..13B}
{Bekki}, K. \& {Couch}, W.~J. 2003, \apjl, 596, L13

\bibitem[{{Biffi} {et~al.}(2018){Biffi}, {Dolag}, \& {Merloni}}]{2018MNRAS.481.2213B}
{Biffi}, V., {Dolag}, K., \& {Merloni}, A. 2018, \mnras, 481, 2213

\bibitem[{{Bondi}(1952)}]{1952MNRAS.112..195B}
{Bondi}, H. 1952, \mnras, 112, 195

\bibitem[{{Boselli} {et~al.}(2016){Boselli}, {Cuillandre}, {Fossati}, {Boissier}, {Bomans}, {Consolandi}, {Anselmi}, {Cortese}, {C{\^o}t{\'e}}, {Durrell}, {Ferrarese}, {Fumagalli}, {Gavazzi}, {Gwyn}, {Hensler}, {Sun}, \& {Toloba}}]{2016A&A...587A..68B}
{Boselli}, A., {Cuillandre}, J.~C., {Fossati}, M., {et~al.} 2016, \aap, 587, A68

\bibitem[{{Boselli} {et~al.}(2022){Boselli}, {Fossati}, \& {Sun}}]{2022A&ARv..30....3B}
{Boselli}, A., {Fossati}, M., \& {Sun}, M. 2022, \aapr, 30, 3

\bibitem[{{Brandt} \& {Hasinger}(2005)}]{2005ARA&A..43..827B}
{Brandt}, W.~N. \& {Hasinger}, G. 2005, \araa, 43, 827

\bibitem[{{Byrd} \& {Valtonen}(1990)}]{1990ApJ...350...89B}
{Byrd}, G. \& {Valtonen}, M. 1990, \apj, 350, 89

\bibitem[{{Cantalupo}(2010)}]{2010MNRAS.403L..16C}
{Cantalupo}, S. 2010, \mnras, 403, L16

\bibitem[{{Choque-Challapa} {et~al.}(2019){Choque-Challapa}, {Smith}, {Candlish}, {Peletier}, \& {Shin}}]{2019MNRAS.490.3654C}
{Choque-Challapa}, N., {Smith}, R., {Candlish}, G., {Peletier}, R., \& {Shin}, J. 2019, \mnras, 490, 3654

\bibitem[{{Cohen} {et~al.}(2014){Cohen}, {Hickox}, {Wegner}, {Einasto}, \& {Vennik}}]{2014ApJ...783..136C}
{Cohen}, S.~A., {Hickox}, R.~C., {Wegner}, G.~A., {Einasto}, M., \& {Vennik}, J. 2014, \apj, 783, 136

\bibitem[{{Di Matteo} {et~al.}(2005){Di Matteo}, {Springel}, \& {Hernquist}}]{2005Natur.433..604D}
{Di Matteo}, T., {Springel}, V., \& {Hernquist}, L. 2005, \nat, 433, 604

\bibitem[{{Diamond-Stanic} \& {Rieke}(2012)}]{2012ApJ...746..168D}
{Diamond-Stanic}, A.~M. \& {Rieke}, G.~H. 2012, \apj, 746, 168

\bibitem[{{Dolag} {et~al.}(2009){Dolag}, {Borgani}, {Murante}, \& {Springel}}]{2009MNRAS.399..497D}
{Dolag}, K., {Borgani}, S., {Murante}, G., \& {Springel}, V. 2009, \mnras, 399, 497

\bibitem[{{Dolag} {et~al.}(2016){Dolag}, {Komatsu}, \& {Sunyaev}}]{2016MNRAS.463.1797D}
{Dolag}, K., {Komatsu}, E., \& {Sunyaev}, R. 2016, \mnras, 463, 1797

\bibitem[{{Dolag} {et~al.}(2005){Dolag}, {Vazza}, {Brunetti}, \& {Tormen}}]{2005MNRAS.364..753D}
{Dolag}, K., {Vazza}, F., {Brunetti}, G., \& {Tormen}, G. 2005, \mnras, 364, 753

\bibitem[{{Ehlert} {et~al.}(2015){Ehlert}, {Allen}, {Brandt}, {Canning}, {Luo}, {Mantz}, {Morris}, {von der Linden}, \& {Xue}}]{2015MNRAS.446.2709E}
{Ehlert}, S., {Allen}, S.~W., {Brandt}, W.~N., {et~al.} 2015, \mnras, 446, 2709

\bibitem[{{Ehlert} {et~al.}(2014){Ehlert}, {von der Linden}, {Allen}, {Brandt}, {Xue}, {Luo}, {Mantz}, {Morris}, {Applegate}, \& {Kelly}}]{2014MNRAS.437.1942E}
{Ehlert}, S., {von der Linden}, A., {Allen}, S.~W., {et~al.} 2014, \mnras, 437, 1942

\bibitem[{{Evrard}(1991)}]{1991MNRAS.248P...8E}
{Evrard}, A.~E. 1991, \mnras, 248, 8P

\bibitem[{{Fabian}(2012)}]{2012ARA&A..50..455F}
{Fabian}, A.~C. 2012, \araa, 50, 455

\bibitem[{{Fabjan} {et~al.}(2010){Fabjan}, {Borgani}, {Tornatore}, {Saro}, {Murante}, \& {Dolag}}]{2010MNRAS.401.1670F}
{Fabjan}, D., {Borgani}, S., {Tornatore}, L., {et~al.} 2010, \mnras, 401, 1670

\bibitem[{{Fassbender} {et~al.}(2012){Fassbender}, {{\v{S}}uhada}, \& {Nastasi}}]{2012AdAst2012E..32F}
{Fassbender}, R., {{\v{S}}uhada}, R., \& {Nastasi}, A. 2012, Advances in Astronomy, 2012, 138380

\bibitem[{{Florez} {et~al.}(2020){Florez}, {Jogee}, {Sherman}, {Stevans}, {Finkelstein}, {Papovich}, {Kawinwanichakij}, {Ciardullo}, {Gronwall}, {Urry}, {Kirkpatrick}, {LaMassa}, {Ananna}, \& {Wold}}]{2020MNRAS.497.3273F}
{Florez}, J., {Jogee}, S., {Sherman}, S., {et~al.} 2020, \mnras, 497, 3273

\bibitem[{{Franx} {et~al.}(2008){Franx}, {van Dokkum}, {F{\"o}rster Schreiber}, {Wuyts}, {Labb{\'e}}, \& {Toft}}]{2008ApJ...688..770F}
{Franx}, M., {van Dokkum}, P.~G., {F{\"o}rster Schreiber}, N.~M., {et~al.} 2008, \apj, 688, 770

\bibitem[{{Gavazzi} {et~al.}(2006){Gavazzi}, {Boselli}, {Cortese}, {Arosio}, {Gallazzi}, {Pedotti}, \& {Carrasco}}]{2006A&A...446..839G}
{Gavazzi}, G., {Boselli}, A., {Cortese}, L., {et~al.} 2006, \aap, 446, 839

\bibitem[{{Gunn} \& {Gott}(1972)}]{1972ApJ...176....1G}
{Gunn}, J.~E. \& {Gott}, J.~Richard, I. 1972, \apj, 176, 1

\bibitem[{{Haardt} \& {Madau}(2001)}]{2001cghr.confE..64H}
{Haardt}, F. \& {Madau}, P. 2001, in Clusters of Galaxies and the High Redshift Universe Observed in X-rays, ed. D.~M. {Neumann} \& J.~T.~V. {Tran}, 64

\bibitem[{{Haggar} {et~al.}(2023){Haggar}, {Kuchner}, {Gray}, {Pearce}, {Knebe}, {Yepes}, \& {Cui}}]{2023MNRAS.518.1316H}
{Haggar}, R., {Kuchner}, U., {Gray}, M.~E., {et~al.} 2023, \mnras, 518, 1316

\bibitem[{{Haggard} {et~al.}(2010){Haggard}, {Green}, {Anderson}, {Constantin}, {Aldcroft}, {Kim}, \& {Barkhouse}}]{2010ApJ...723.1447H}
{Haggard}, D., {Green}, P.~J., {Anderson}, S.~F., {et~al.} 2010, \apj, 723, 1447

\bibitem[{{Haines} {et~al.}(2012){Haines}, {Pereira}, {Sanderson}, {Smith}, {Egami}, {Babul}, {Edge}, {Finoguenov}, {Moran}, \& {Okabe}}]{2012ApJ...754...97H}
{Haines}, C.~P., {Pereira}, M.~J., {Sanderson}, A.~J.~R., {et~al.} 2012, \apj, 754, 97

\bibitem[{{Hernquist} \& {Mihos}(1995)}]{1995ApJ...448...41H}
{Hernquist}, L. \& {Mihos}, J.~C. 1995, \apj, 448, 41

\bibitem[{{Hirschmann} {et~al.}(2014){Hirschmann}, {Dolag}, {Saro}, {Bachmann}, {Borgani}, \& {Burkert}}]{2014MNRAS.442.2304H}
{Hirschmann}, M., {Dolag}, K., {Saro}, A., {et~al.} 2014, \mnras, 442, 2304

\bibitem[{{Hopkins} {et~al.}(2008){Hopkins}, {Hernquist}, {Cox}, \& {Kere{\v{s}}}}]{2008ApJS..175..356H}
{Hopkins}, P.~F., {Hernquist}, L., {Cox}, T.~J., \& {Kere{\v{s}}}, D. 2008, \apjs, 175, 356

\bibitem[{{Huertas-Company} {et~al.}(2016){Huertas-Company}, {Bernardi}, {P{\'e}rez-Gonz{\'a}lez}, {Ashby}, {Barro}, {Conselice}, {Daddi}, {Dekel}, {Dimauro}, {Faber}, {Grogin}, {Kartaltepe}, {Kocevski}, {Koekemoer}, {Koo}, {Mei}, \& {Shankar}}]{2016MNRAS.462.4495H}
{Huertas-Company}, M., {Bernardi}, M., {P{\'e}rez-Gonz{\'a}lez}, P.~G., {et~al.} 2016, \mnras, 462, 4495

\bibitem[{{Hwang} {et~al.}(2019){Hwang}, {Shin}, \& {Song}}]{2019MNRAS.489..339H}
{Hwang}, H.~S., {Shin}, J., \& {Song}, H. 2019, \mnras, 489, 339

\bibitem[{{Iannuzzi} \& {Dolag}(2012)}]{2012MNRAS.427.1024I}
{Iannuzzi}, F. \& {Dolag}, K. 2012, \mnras, 427, 1024

\bibitem[{{Kapferer} {et~al.}(2008){Kapferer}, {Kronberger}, {Ferrari}, {Riser}, \& {Schindler}}]{2008MNRAS.389.1405K}
{Kapferer}, W., {Kronberger}, T., {Ferrari}, C., {Riser}, T., \& {Schindler}, S. 2008, \mnras, 389, 1405

\bibitem[{{Kapferer} {et~al.}(2009){Kapferer}, {Sluka}, {Schindler}, {Ferrari}, \& {Ziegler}}]{2009A&A...499...87K}
{Kapferer}, W., {Sluka}, C., {Schindler}, S., {Ferrari}, C., \& {Ziegler}, B. 2009, \aap, 499, 87

\bibitem[{{Kennicutt}(1983)}]{1983AJ.....88..483K}
{Kennicutt}, R.~C., J. 1983, \aj, 88, 483

\bibitem[{{King} \& {Pounds}(2015)}]{2015ARA&A..53..115K}
{King}, A. \& {Pounds}, K. 2015, \araa, 53, 115

\bibitem[{{Komatsu} {et~al.}(2009){Komatsu}, {Dunkley}, {Nolta}, {Bennett}, {Gold}, {Hinshaw}, {Jarosik}, {Larson}, {Limon}, {Page}, {Spergel}, {Halpern}, {Hill}, {Kogut}, {Meyer}, {Tucker}, {Weiland}, {Wollack}, \& {Wright}}]{2009ApJS..180..330K}
{Komatsu}, E., {Dunkley}, J., {Nolta}, M.~R., {et~al.} 2009, \apjs, 180, 330

\bibitem[{{Koulouridis} \& {Bartalucci}(2019)}]{2019A&A...623L..10K}
{Koulouridis}, E. \& {Bartalucci}, I. 2019, \aap, 623, L10

\bibitem[{{Koulouridis} \& {Plionis}(2010)}]{2010ApJ...714L.181K}
{Koulouridis}, E. \& {Plionis}, M. 2010, \apjl, 714, L181

\bibitem[{{Koulouridis} {et~al.}(2018){Koulouridis}, {Ricci}, {Giles}, {Adami}, {Ramos-Ceja}, {Pierre}, {Plionis}, {Lidman}, {Georgantopoulos}, {Chiappetti}, {Elyiv}, {Ettori}, {Faccioli}, {Fotopoulou}, {Gastaldello}, {Pacaud}, {Paltani}, \& {Vignali}}]{2018A&A...620A..20K}
{Koulouridis}, E., {Ricci}, M., {Giles}, P., {et~al.} 2018, \aap, 620, A20

\bibitem[{{Kronberger} {et~al.}(2008){Kronberger}, {Kapferer}, {Ferrari}, {Unterguggenberger}, \& {Schindler}}]{2008A&A...481..337K}
{Kronberger}, T., {Kapferer}, W., {Ferrari}, C., {Unterguggenberger}, S., \& {Schindler}, S. 2008, \aap, 481, 337

\bibitem[{{Larson} {et~al.}(1980){Larson}, {Tinsley}, \& {Caldwell}}]{1980ApJ...237..692L}
{Larson}, R.~B., {Tinsley}, B.~M., \& {Caldwell}, C.~N. 1980, \apj, 237, 692

\bibitem[{{Leslie} {et~al.}(2016){Leslie}, {Kewley}, {Sanders}, \& {Lee}}]{2016MNRAS.455L..82L}
{Leslie}, S.~K., {Kewley}, L.~J., {Sanders}, D.~B., \& {Lee}, N. 2016, \mnras, 455, L82

\bibitem[{{Liu} {et~al.}(2019){Liu}, {Hao}, {Wang}, \& {Yang}}]{2019ApJ...878...69L}
{Liu}, C., {Hao}, L., {Wang}, H., \& {Yang}, X. 2019, \apj, 878, 69

\bibitem[{{Lotz} {et~al.}(2019){Lotz}, {Remus}, {Dolag}, {Biviano}, \& {Burkert}}]{2019MNRAS.488.5370L}
{Lotz}, M., {Remus}, R.-S., {Dolag}, K., {Biviano}, A., \& {Burkert}, A. 2019, \mnras, 488, 5370

\bibitem[{{Magorrian} {et~al.}(1998){Magorrian}, {Tremaine}, {Richstone}, {Bender}, {Bower}, {Dressler}, {Faber}, {Gebhardt}, {Green}, {Grillmair}, {Kormendy}, \& {Lauer}}]{1998AJ....115.2285M}
{Magorrian}, J., {Tremaine}, S., {Richstone}, D., {et~al.} 1998, \aj, 115, 2285

\bibitem[{{Maio} {et~al.}(2013){Maio}, {Dotti}, {Petkova}, {Perego}, \& {Volonteri}}]{2013ApJ...767...37M}
{Maio}, U., {Dotti}, M., {Petkova}, M., {Perego}, A., \& {Volonteri}, M. 2013, \apj, 767, 37

\bibitem[{{Marconi} {et~al.}(2004){Marconi}, {Risaliti}, {Gilli}, {Hunt}, {Maiolino}, \& {Salvati}}]{2004MNRAS.351..169M}
{Marconi}, A., {Risaliti}, G., {Gilli}, R., {et~al.} 2004, \mnras, 351, 169

\bibitem[{{Martini} {et~al.}(2007){Martini}, {Mulchaey}, \& {Kelson}}]{2007ApJ...664..761M}
{Martini}, P., {Mulchaey}, J.~S., \& {Kelson}, D.~D. 2007, \apj, 664, 761

\bibitem[{{Merritt}(1983)}]{1983ApJ...264...24M}
{Merritt}, D. 1983, \apj, 264, 24

\bibitem[{{Montero-Dorta} {et~al.}(2023){Montero-Dorta}, {Rodriguez}, {Artale}, {Smith}, \& {Chaves-Montero}}]{2023MNRAS.tmp.3185M}
{Montero-Dorta}, A.~D., {Rodriguez}, F., {Artale}, M.~C., {Smith}, R., \& {Chaves-Montero}, J. 2023, \mnras [\eprint[arXiv]{2212.12090}]

\bibitem[{{Moore} {et~al.}(1996){Moore}, {Katz}, {Lake}, {Dressler}, \& {Oemler}}]{1996Natur.379..613M}
{Moore}, B., {Katz}, N., {Lake}, G., {Dressler}, A., \& {Oemler}, A. 1996, \nat, 379, 613

\bibitem[{{Moore} {et~al.}(1998){Moore}, {Lake}, \& {Katz}}]{1998ApJ...495..139M}
{Moore}, B., {Lake}, G., \& {Katz}, N. 1998, \apj, 495, 139

\bibitem[{{Mullaney} {et~al.}(2012){Mullaney}, {Daddi}, {B{\'e}thermin}, {Elbaz}, {Juneau}, {Pannella}, {Sargent}, {Alexander}, \& {Hickox}}]{2012ApJ...753L..30M}
{Mullaney}, J.~R., {Daddi}, E., {B{\'e}thermin}, M., {et~al.} 2012, \apjl, 753, L30

\bibitem[{{Muzzin} {et~al.}(2012){Muzzin}, {Wilson}, {Yee}, {Gilbank}, {Hoekstra}, {Demarco}, {Balogh}, {van Dokkum}, {Franx}, {Ellingson}, {Hicks}, {Nantais}, {Noble}, {Lacy}, {Lidman}, {Rettura}, {Surace}, \& {Webb}}]{2012ApJ...746..188M}
{Muzzin}, A., {Wilson}, G., {Yee}, H.~K.~C., {et~al.} 2012, \apj, 746, 188

\bibitem[{{Nandra} {et~al.}(2007){Nandra}, {Georgakakis}, {Willmer}, {Cooper}, {Croton}, {Davis}, {Faber}, {Koo}, {Laird}, \& {Newman}}]{2007ApJ...660L..11N}
{Nandra}, K., {Georgakakis}, A., {Willmer}, C.~N.~A., {et~al.} 2007, \apjl, 660, L11

\bibitem[{{Navarro} {et~al.}(1995){Navarro}, {Frenk}, \& {White}}]{1995MNRAS.275..720N}
{Navarro}, J.~F., {Frenk}, C.~S., \& {White}, S. D.~M. 1995, \mnras, 275, 720

\bibitem[{{Navarro} {et~al.}(1996){Navarro}, {Frenk}, \& {White}}]{1996ApJ...462..563N}
{Navarro}, J.~F., {Frenk}, C.~S., \& {White}, S. D.~M. 1996, \apj, 462, 563

\bibitem[{{Navarro} {et~al.}(1997){Navarro}, {Frenk}, \& {White}}]{1997ApJ...490..493N}
{Navarro}, J.~F., {Frenk}, C.~S., \& {White}, S. D.~M. 1997, \apj, 490, 493

\bibitem[{{Park} {et~al.}(2006){Park}, {Kashyap}, {Siemiginowska}, {van Dyk}, {Zezas}, {Heinke}, \& {Wargelin}}]{2006ApJ...652..610P}
{Park}, T., {Kashyap}, V.~L., {Siemiginowska}, A., {et~al.} 2006, \apj, 652, 610

\bibitem[{{Pe{\~n}arrubia} {et~al.}(2008){Pe{\~n}arrubia}, {Navarro}, \& {McConnachie}}]{2008ApJ...673..226P}
{Pe{\~n}arrubia}, J., {Navarro}, J.~F., \& {McConnachie}, A.~W. 2008, \apj, 673, 226

\bibitem[{{Peng} {et~al.}(2015){Peng}, {Maiolino}, \& {Cochrane}}]{2015Natur.521..192P}
{Peng}, Y., {Maiolino}, R., \& {Cochrane}, R. 2015, \nat, 521, 192

\bibitem[{{Peng} {et~al.}(2010){Peng}, {Lilly}, {Kova{\v{c}}}, {Bolzonella}, {Pozzetti}, {Renzini}, {Zamorani}, {Ilbert}, {Knobel}, {Iovino}, {Maier}, {Cucciati}, {Tasca}, {Carollo}, {Silverman}, {Kampczyk}, {de Ravel}, {Sanders}, {Scoville}, {Contini}, {Mainieri}, {Scodeggio}, {Kneib}, {Le F{\`e}vre}, {Bardelli}, {Bongiorno}, {Caputi}, {Coppa}, {de la Torre}, {Franzetti}, {Garilli}, {Lamareille}, {Le Borgne}, {Le Brun}, {Mignoli}, {Perez Montero}, {Pello}, {Ricciardelli}, {Tanaka}, {Tresse}, {Vergani}, {Welikala}, {Zucca}, {Oesch}, {Abbas}, {Barnes}, {Bordoloi}, {Bottini}, {Cappi}, {Cassata}, {Cimatti}, {Fumana}, {Hasinger}, {Koekemoer}, {Leauthaud}, {Maccagni}, {Marinoni}, {McCracken}, {Memeo}, {Meneux}, {Nair}, {Porciani}, {Presotto}, \& {Scaramella}}]{2010ApJ...721..193P}
{Peng}, Y.-j., {Lilly}, S.~J., {Kova{\v{c}}}, K., {et~al.} 2010, \apj, 721, 193

\bibitem[{{Poggianti} {et~al.}(2021){Poggianti}, {Bellhouse}, {Deb}, {Franchetto}, {Fritz}, {George}, {Gullieuszik}, {Jaff{\'e}}, {Moretti}, {Mueller}, {Radovich}, {Ramatsoku}, {Vulcani}, \& {GASP Team}}]{2021IAUS..359..108P}
{Poggianti}, B.~M., {Bellhouse}, C., {Deb}, T., {et~al.} 2021, in Galaxy Evolution and Feedback across Different Environments, ed. T.~{Storchi Bergmann}, W.~{Forman}, R.~{Overzier}, \& R.~{Riffel}, Vol. 359, 108--116

\bibitem[{{Poggianti} {et~al.}(2017){Poggianti}, {Jaff{\'e}}, {Moretti}, {Gullieuszik}, {Radovich}, {Tonnesen}, {Fritz}, {Bettoni}, {Vulcani}, {Fasano}, {Bellhouse}, {Hau}, \& {Omizzolo}}]{2017Natur.548..304P}
{Poggianti}, B.~M., {Jaff{\'e}}, Y.~L., {Moretti}, A., {et~al.} 2017, \nat, 548, 304

\bibitem[{{Quilis} {et~al.}(2000){Quilis}, {Moore}, \& {Bower}}]{2000Sci...288.1617Q}
{Quilis}, V., {Moore}, B., \& {Bower}, R. 2000, Science, 288, 1617

\bibitem[{{Read} {et~al.}(2006){Read}, {Wilkinson}, {Evans}, {Gilmore}, \& {Kleyna}}]{2006MNRAS.366..429R}
{Read}, J.~I., {Wilkinson}, M.~I., {Evans}, N.~W., {Gilmore}, G., \& {Kleyna}, J.~T. 2006, \mnras, 366, 429

\bibitem[{{Ruderman} \& {Ebeling}(2005)}]{2005ApJ...623L..81R}
{Ruderman}, J.~T. \& {Ebeling}, H. 2005, \apjl, 623, L81

\bibitem[{{Sanders} {et~al.}(1988){Sanders}, {Soifer}, {Elias}, {Madore}, {Matthews}, {Neugebauer}, \& {Scoville}}]{1988ApJ...325...74S}
{Sanders}, D.~B., {Soifer}, B.~T., {Elias}, J.~H., {et~al.} 1988, \apj, 325, 74

\bibitem[{{Saro} {et~al.}(2010){Saro}, {De Lucia}, {Borgani}, \& {Dolag}}]{2010MNRAS.406..729S}
{Saro}, A., {De Lucia}, G., {Borgani}, S., \& {Dolag}, K. 2010, \mnras, 406, 729

\bibitem[{{Smethurst} {et~al.}(2017){Smethurst}, {Lintott}, {Bamford}, {Hart}, {Kruk}, {Masters}, {Nichol}, \& {Simmons}}]{2017MNRAS.469.3670S}
{Smethurst}, R.~J., {Lintott}, C.~J., {Bamford}, S.~P., {et~al.} 2017, \mnras, 469, 3670

\bibitem[{{Smith} {et~al.}(2016){Smith}, {Choi}, {Lee}, {Rhee}, {Sanchez-Janssen}, \& {Yi}}]{2016ApJ...833..109S}
{Smith}, R., {Choi}, H., {Lee}, J., {et~al.} 2016, \apj, 833, 109

\bibitem[{{Smith} {et~al.}(2013){Smith}, {S{\'a}nchez-Janssen}, {Fellhauer}, {Puzia}, {Aguerri}, \& {Farias}}]{2013MNRAS.429.1066S}
{Smith}, R., {S{\'a}nchez-Janssen}, R., {Fellhauer}, M., {et~al.} 2013, \mnras, 429, 1066

\bibitem[{{Springel}(2005)}]{2005MNRAS.364.1105S}
{Springel}, V. 2005, \mnras, 364, 1105

\bibitem[{{Springel} {et~al.}(2005{\natexlab{a}}){Springel}, {Di Matteo}, \& {Hernquist}}]{2005MNRAS.361..776S}
{Springel}, V., {Di Matteo}, T., \& {Hernquist}, L. 2005{\natexlab{a}}, \mnras, 361, 776

\bibitem[{{Springel} \& {Hernquist}(2003)}]{2003MNRAS.339..289S}
{Springel}, V. \& {Hernquist}, L. 2003, \mnras, 339, 289

\bibitem[{{Springel} {et~al.}(2005{\natexlab{b}}){Springel}, {White}, {Jenkins}, {Frenk}, {Yoshida}, {Gao}, {Navarro}, {Thacker}, {Croton}, {Helly}, {Peacock}, {Cole}, {Thomas}, {Couchman}, {Evrard}, {Colberg}, \& {Pearce}}]{2005Natur.435..629S}
{Springel}, V., {White}, S. D.~M., {Jenkins}, A., {et~al.} 2005{\natexlab{b}}, \nat, 435, 629

\bibitem[{{Springel} {et~al.}(2001){Springel}, {White}, {Tormen}, \& {Kauffmann}}]{2001MNRAS.328..726S}
{Springel}, V., {White}, S. D.~M., {Tormen}, G., \& {Kauffmann}, G. 2001, \mnras, 328, 726

\bibitem[{{Steinborn} {et~al.}(2015){Steinborn}, {Dolag}, {Hirschmann}, {Prieto}, \& {Remus}}]{2015MNRAS.448.1504S}
{Steinborn}, L.~K., {Dolag}, K., {Hirschmann}, M., {Prieto}, M.~A., \& {Remus}, R.-S. 2015, \mnras, 448, 1504

\bibitem[{{Steinborn} {et~al.}(2018){Steinborn}, {Hirschmann}, {Dolag}, {Shankar}, {Juneau}, {Krumpe}, {Remus}, \& {Teklu}}]{2018MNRAS.481..341S}
{Steinborn}, L.~K., {Hirschmann}, M., {Dolag}, K., {et~al.} 2018, \mnras, 481, 341

\bibitem[{{Teklu} {et~al.}(2017){Teklu}, {Remus}, {Dolag}, \& {Burkert}}]{2017MNRAS.472.4769T}
{Teklu}, A.~F., {Remus}, R.-S., {Dolag}, K., \& {Burkert}, A. 2017, \mnras, 472, 4769

\bibitem[{{Tonnesen} \& {Bryan}(2009)}]{2009ApJ...694..789T}
{Tonnesen}, S. \& {Bryan}, G.~L. 2009, \apj, 694, 789

\bibitem[{{Tornatore} {et~al.}(2007){Tornatore}, {Borgani}, {Dolag}, \& {Matteucci}}]{2007MNRAS.382.1050T}
{Tornatore}, L., {Borgani}, S., {Dolag}, K., \& {Matteucci}, F. 2007, \mnras, 382, 1050

\bibitem[{{Tornatore} {et~al.}(2004){Tornatore}, {Borgani}, {Matteucci}, {Recchi}, \& {Tozzi}}]{2004MNRAS.349L..19T}
{Tornatore}, L., {Borgani}, S., {Matteucci}, F., {Recchi}, S., \& {Tozzi}, P. 2004, \mnras, 349, L19

\bibitem[{{von der Linden} {et~al.}(2010){von der Linden}, {Wild}, {Kauffmann}, {White}, \& {Weinmann}}]{2010MNRAS.404.1231V}
{von der Linden}, A., {Wild}, V., {Kauffmann}, G., {White}, S. D.~M., \& {Weinmann}, S. 2010, \mnras, 404, 1231

\bibitem[{{Vulcani} {et~al.}(2023){Vulcani}, {Poggianti}, {Gullieuszik}, {Moretti}, {Fritz}, {Bettoni}, {Facciolli}, {Fasano}, \& {Omizzolo}}]{2023arXiv230202376V}
{Vulcani}, B., {Poggianti}, B.~M., {Gullieuszik}, M., {et~al.} 2023, arXiv e-prints, arXiv:2302.02376

\bibitem[{{Wetzel} {et~al.}(2012){Wetzel}, {Tinker}, \& {Conroy}}]{2012MNRAS.424..232W}
{Wetzel}, A.~R., {Tinker}, J.~L., \& {Conroy}, C. 2012, \mnras, 424, 232

\bibitem[{{Wiersma} {et~al.}(2009){Wiersma}, {Schaye}, \& {Smith}}]{2009MNRAS.393...99W}
{Wiersma}, R. P.~C., {Schaye}, J., \& {Smith}, B.~D. 2009, \mnras, 393, 99

\end{thebibliography}
%

\end{document}